\documentclass[screen,sigplan]{acmart}
\renewcommand\footnotetextcopyrightpermission[1]{}

\usepackage{tikz}
\usepackage{amsmath}

\usepackage{filecontents}
\usepackage{balance}
\usepackage[normalem]{ulem}
\usepackage{breakurl}
\usepackage{url}
\usepackage{graphicx}
\usepackage{subfig}
\usepackage{booktabs}
\usepackage{svg}
\usepackage{thmbox}
\usepackage{pifont}% http://ctan.org/pkg/pifont
\usepackage{bbding}
\usepackage{fancyhdr}
\usepackage{xspace}
\usepackage{mfirstuc} % for cap the first letter for a cmd
\usepackage{hyperref}
\usepackage{multirow}
\usepackage{wrapfig}

\usepackage[labelfont=bf,font=normal,skip=1pt]{caption}

\usepackage{titlesec}
\titlespacing*\section{3pt}{3pt plus 2pt minus 1pt}{3pt plus 2pt minus 1pt}
\titlespacing*\subsection{2pt}{3pt plus 2pt minus 1pt}{3pt plus 2pt minus 1pt}

\newcommand{\papercomment}[2]{{\bf[\textcolor{red}{#1}: \textcolor{blue}{#2}]}}
\newcommand{\TODO}[1]{{\papercomment{TODO}{#1}}}

\newcommand{\cheng}[1]{{\papercomment{cheng}{#1}}}
\newcommand{\yifan}[1]{{\papercomment{yifan}{#1}}}

\def\hn{\sffamily\selectfont}
\newcommand{\mpfont}{\hn\scriptsize}
\newcommand{\MPworker}[2]{\unskip{\color{#1}\vrule\vrule}{\marginpar{\raggedright\color{#1}\mpfont #2}}}
\newcommand{\CP}[1]{\MPworker{blue}{CT: #1}}

\newcommand{\sys}{PipeLLM\xspace} % incorporate training
\let\latexusecounter=\usecounter
  {\begin{list}{\labelitemi}{\itemsep3pt \topsep3pt \parsep0.00in
  \partopsep=3pt \leftmargin1em}}%
  {\end{list}}
\newenvironment{myitemize2}%
  {\begin{list}{\labelitemi}{\itemsep1pt \topsep2pt \parsep0.00in
  \partopsep=0pt \leftmargin1.2em}}%
  {\end{list}}

\def\compactsortof{\itemsep=0in \topsep=2pt \parsep=0.00in \partopsep=0pt \leftmargin=1.2em}
\newenvironment{myenumerate2}
  {\def\usecounter{\compactsortof\latexusecounter}
   \begin{enumerate}}
  {\end{enumerate}\let\usecounter=\latexusecounter}

\newcommand{\CF}[1]{\xmakefirstuc{#1}}
\newcommand{\heading}[1]{
  \vspace{1ex}
  \noindent
  \textbf{#1}}

\newcommand{\kvcache}{KV cache\xspace}
\newcommand{\techone}{speculative pipelined encryption\xspace}

\newcommand{\cc}{confidential computing\xspace}
\newcommand{\ccshort}{CC\xspace}
\newcommand{\ccenabled}{CC-enabled\xspace}
\newcommand{\ccdisabled}{CC-disabled\xspace}

\newcommand{\nv}{NVIDIA\xspace}

\newcommand{\nvcc}{NVIDIA Confidential Computing\xspace}

\newcommand{\flexgen}{FlexGen\xspace}
\newcommand{\vllm}{vLLM\xspace}
\newcommand{\peft}{PEFT\xspace}

\begin{document}

%don't want date printed
\date{}

\title{{\sys}: Fast and Confidential Large Language Model Services with Speculative Pipelined Encryption}

%for single author (just remove % characters)
\author{
\rm Yifan Tan$^{1,3}$,\; Cheng Tan$^{2}$,\; Zeyu Mi$^{1,3}$,\; and Haibo Chen$^{1,3}$\\
{\normalsize {$^1$Institute of Parallel and Distributed Systems, SEIEE, Shanghai Jiao Tong University}} \\
{\normalsize {$^2$Northeastern University}}\\
{\normalsize {$^3$Engineering Research Center for Domain-specific Operating Systems, Ministry of Education, China}}
} % end author

\subtitle{\textit{To appear in ASPLOS 2025}}

\begin{abstract}

%Cloud-based Large Language Models (LLMs) promise performance elasticity, high availability, and ease of deployment. %, etc.
%However, such an approach exposes both the models and user requests to potential security risks within an untrusted cloud.
%Confidential computing on GPUs, such as NVIDIA's H100,
%presents a system-level approach to this issue by implementing data encryption and strong isolation.
%
Confidential computing on GPUs, like NVIDIA H100,
mitigates the security risks of outsourced Large Language Models (LLMs) 
by implementing strong isolation and data encryption.
Nonetheless, this encryption incurs a significant performance overhead,
reaching up to 52.8\% and 88.2\% throughput drop when serving OPT-30B and OPT-66B,
respectively.
% [cheng: the overhead is vastly different for large and mid-sized models, we
% might give two numbers here]

To address this challenge, we introduce \emph{\sys},
a user-transparent runtime system.
%inspired by the CPU instruction pipelining technique. [secondary info]
{\sys} removes the overhead %nearly all overhead %[cheng: hard to claim this with 19% overhead]
by overlapping the encryption and GPU computation %\CP{Yifan, check} \yifan{ok}
%\yifan{overlap encryption with computation}
through pipelining---an idea inspired by the CPU instruction pipelining---thereby
effectively concealing the latency increase caused by encryption.
The primary technical challenge is that, unlike CPUs,
the encryption module lacks prior knowledge of the specific data needing encryption until it is requested by the GPUs.
To this end, we propose \emph{\techone} to predict the data requiring encryption by analyzing the serving patterns of LLMs.
Further, we have developed an efficient, low-cost pipeline relinquishing approach for instances of incorrect predictions.
Our experiments on NVIDIA H100 GPU show that
compared with vanilla systems without confidential computing (e.g., vLLM, PEFT, and FlexGen),
% [cheng: Yifan, let's use Oxford comma, meaning "A, B, and C"---notice the comma before the "and"]
{\sys} incurs modest overhead ($<$19.6\% in throughput) across various LLM sizes, from 13B to 175B.

\iffalse
%Large language models (LLMs) have wide applications.
%  [cheng: this is blah]
%
Serving LLMs on cloud enjoys many benefits
including cloud's scalability, availability, and performance,
but it expose models and user requests to the cloud.
%increases the risk of data leak.
%
GPU hardware confidential computing (like H100) provides a system-level
approach to address the problem by encrypting the data.
%against attackers.
The encryption however brings up to 33\% performance overhead in LLM serving.
%
To mitigate this problem, we propose {\sys},
a user-transparent runtime inspired by the CPU instruction pipelining.
% [cheng: rethink: do we want to emphasize this?]
%mechanism in CPU micro-architecture.
%
It removes (almost all) overheads
by overlapping the encryption and data transmission using pipelining,
effectively hiding the encryption overheads.
%
The technical challenge is that, unlike CPUs,
the encryption modules do not know what to encrypt
until GPUs require the data.
%
To conquer the challenge,
we introduce \emph{\techone} to predict the to-be-encrypted data
by leveraging LLMs' serving patterns.
We also design a low-cost pipeline relinquishing approach
for incorrect predictions.
%
%the pipeline flush is rare because of the LLM pattern.
%
Our experiments show that {\sys}'s overhead is $<$6\%
in applications.
\fi
\end{abstract}

\maketitle

\pagestyle{plain}

\section{Introduction}

% \cheng{TODO:
% - need a new name; we're no longer focus on inference
% - TBD: are we still only focusing on LLMs? (I vote so)
% }

%- motivation: confidential LLM
%Large Language Models (LLMs) have exhibited exceptional capabilities in understanding and generating human languages,
%as well as in handling complex code and textual documents.
%Due to these capabilities, they are being increasingly used in a variety of
%applications, including chatbots~\cite{zheng2023judging} and pair programming
%tools~\cite{copilot}.
%With the expansion of open-source LLMs~\cite{touvron2023llama,jiang2024mixtral,zhang2022opt},
%an increasing number of companies are poised to integrate their own models,
%which are fine-tuned based on open-source LLMs, into their business operations.\CP{can be significantly cut}

Large Language Models (LLMs) are increasingly used across various
applications~\cite{zheng2023judging,copilot}.
With the growth of open-source
LLMs~\cite{touvron2023llama,jiang2024mixtral,zhang2022opt}, companies are
integrating and fine-tuning these models into their business operations.

Due to LLM's reliance on high-end GPUs, many businesses opt for
outsourced services, such as cloud,
attracted by their high availability and flexible pay-as-you-go models.
However, these cloud infrastructures, often complex in nature, encompass a large Trusted Computing Base (TCB),
which may %is known to [cheng: this is too strong]
contain vulnerabilities, both publicly reported and zero-day~\cite{li2021twinvisor, chen2023security}.
This poses security risks for LLMs, which are usually fine-tuned with proprietary data,
and %the \yifan{remove ``the''?} 
user prompts that contain sensitive business information.
%The deployed models are usually fine-tuned with a wealth of proprietary,
%domain-specific data derived from business operations. Any unauthorized disclosure of these models leads to significant financial losses.
%In addition, the process of serving LLMs involves handling prompts that contain sensitive business information.
Thus, any data breach could expose critical business secrets.

%- confidential computing as a solution
%  -- software CC
%  -- HW CC

%- a parallel: CPU confidential computing
%  -- we believe HW CC could be the production use

%- HW CC intro
%  - H100 is the first GPU with CC
%  - more are coming (are there any?)

To mitigate these security threats,
people introduce confidential computing.
Confidential computing is designed to safeguard tenants' code and data against
untrusted privileged software and %potentially inquisitive cloud operators.
rogue employees of cloud providers.
The confidential virtual machine (CVM),
supported by technologies such as Intel TDX~\cite{tdxmodule}, AMD SEV~\cite{sevsnp}, and ARM CCA~\cite{armcca},
serves as a prime example of this.
Any software external to a CVM is unable to access the code and data within it.

Regarding machine learning workloads,
people develop GPU enclaves to enhance security measures within GPUs~\cite{volos2018graviton,h100cc}.
A notable implementation of this is the NVIDIA H100 GPU~\cite{h100cc},
which supports confidential computing inside the GPU to protect sensitive data and models from unauthorized access.
Moreover, the data communication between the CVM and the GPU enclave is encrypted, further reinforcing the security of I/O operations.

%- problem: HW CC doesn't work well for LLM
%  why? LLM is huge =>
%            GPU memory is limited =>
%              swapping memory between GPU and CPU +
%              IO needs encrypted =>
%                encryption/decryption on CPU becomes bottleneck
%  result: high-overhead for LLM inference

%- figure: w/o CC vs. w/ CC on several cases

Although GPU confidential computing effectively enhances security for traditional small-scale AI models,
it significantly undermines the performance of LLMs in throughput and latency.
%with extensive memory footprints,
%particularly in terms of inference latency and throughput considered in this paper.
Our comprehensive experiments on NVIDIA H100 GPUs reveal that the GPU enclave can incur up to
a 52.8\% latency overhead on serving OPT-30B,
a 36.2\% throughput drop on fine-tuning OPT-30B,
and an 88.2\%
%\CP{Yifan, is this the number of 88.2\% througput drop?}\yifan{Yes, it refers to 88.2\%}
throughput drop on serving OPT-66B (\S\ref{s:bottleneck}).
This overhead is largely due to a combination of
memory swapping plus encryption.
The swapping happens because LLMs consume a huge amount of GPU memory.
%which struggles to accommodate both the static LLM parameters and the dynamic runtime states.
For example,
the OPT-66B model needs approximately 132GB of memory to store all its parameters,
%far \yifan{remove ``far''?}
surpassing the 80GB memory of H100 GPUs.
%High Bandwidth Memory (HBM) of the NVIDIA H100 GPU. % [cheng: HBM is irrelevant here]
Moreover, runtime states such as the Key-Value cache (\kvcache)~\cite{pope2022kv} during LLM inferences
and activation during LLM training
also consume significant GPU memory.

% \yifan{\kvcache usages change with input configuration.}

Owing to the limited GPU memory,
a GPU enclave has to dynamically swap out inactive parameters and/or runtime states to the main memory.
This process requires encrypting the data transferred out of the GPU enclave.
Correspondingly, %on the Confidential VM (CVM),
the CPU cores must decrypt data received from the GPU,
and re-encrypt it before sending it back to the GPU (\S\ref{subsec:bg-h100}).
However, the encryption and decryption pose a severe bottleneck %in the data swapping mechanism
due to the limited computational capability. %of CPUs.
This bottleneck significantly harms the overall performance,
particularly in the context of LLMs.

This paper introduces {\sys}, a system designed to eliminate the performance
overhead associated with GPU confidential computing for LLMs.
Importantly, {\sys} achieves this without requiring any changes to the existing LLM systems or the hardware,
while still upholding the same level of security.
% \yifan{Since we introduce side channel, should we have a weaker claim?}.
% [cheng: this claim is fine since both NVIDIA and us cannot handle side
% channels, meaning neither are secure under side channel attacks]
The underlying principle of {\sys} is straightforward yet effective:
it decouples encryption tasks from the critical path of the memory swapping mechanism,
by leveraging \emph{speculative pipelined encryption} (\S\ref{ss:techone}), a technique we proposed.
Drawing inspiration from the concept of speculative execution in CPUs,
{\sys} anticipates which data blocks will be required by the GPU and pre-encrypts them.
By doing so, {\sys} significantly reduces the overhead of the GPU \cc
by integrating predictions, encryptions, and data transfers into a pipeline.

%- our aim is to reduce the overhead of confidential LLM inference
%  Meanwhile, to be readily used, the system requires the following:
%   -- no modification to the application (LLM serving software)
%   -- no modification to the hardware
%   -- do not substantially increase the CPU cores [revise]
%      otherwise, other services on the same machine can be affected

%- our idea is simple: we remove enc/dec from the critical path
%  by "speculative pipelined encryption".
%    - a swapping technique to encrypt the needed pages a priori
%    - it hides the encryption overhead with the data transfer, a classic technique using in GPU

%- a workflow fig
%  -- vanilla: demonstrates the enc/dec is on the critical path
%  -- ours: pipelined the stages

However, akin to CPU pipelining,
an incorrect prediction could not only waste an individual pre-encrypted data
but also invalidate the entire pipeline of subsequent pre-encrypted data.
% \yifan{This is partially incorrect. The pre-encrypted data could be used with re-ordering in some cases.
% Could we say ``could result in''?}
This is a consequence of the encryption scheme used by GPU enclaves,
designed to prevent replay attacks~\cite{replayattack}.
In the H100's confidential computing,
data
is encrypted using a private key in conjunction with a unique integer known as the \emph{Initialization Vector (IV)}.
The IV is synchronized between the CPU and GPU,
and increments by one with each encryption.
Consequently, if an incorrect piece of data is encrypted in a pipeline,
all subsequent IVs in the pipeline could become invalid,
%necessitating 
requiring re-encryption of the subsequent data with the correct IVs.
We will elaborate on the encryption mechanism and IVs in \S\ref{ss:errorhandler}.

% [cheng: Yifan, is the encryption unit (one need an IV) a page? or it can be
% an arbitrary sized data?]
\newcommand{\pages}{\TODO{pages}}
\newcommand{\page}{\TODO{page}}

To address the challenge,
we observe that LLM systems are highly predictable in swapping, %exhibit a high degree of predictability,
allowing {\sys} to accurately determine the sequence of data being swapped using heuristics.
For instance, FlexGen~\cite{flexgen-paper} and PEFT~\cite{peft} compute the LLM through a layer-by-layer process,
enabling {\sys} to efficiently swap in layer parameters in their respective order.
%\TODO{suggested deletion from reviewers}
%\CP{Yifan, can you be specific? like which reviewer and what the reviewer says}
%\yifan{Reviewer C: computes => compute. Fixed here}
Similarly, %other LLM serving systems such as 
vLLM~\cite{vllm-paper}
uses simple swapping policies like FIFO (First-In, First-Out) and LIFO (Last-In, First-Out).
{\sys} can recognize these swapping policies and use them to predict the future swapping sequence.

Moreover, \sys incorporates several techniques to accelerate the pipeline
and mitigate the cost of prediction errors.
First, \sys develops an efficient validation scheme to verify the correctness
of a pre-defined ciphertext (\S\ref{ss:s5-validation}).
Second, \sys introduces \emph{request re-ordering} and \emph{NOP padding} to handle IV mismatches
without relinquishing the entire pipeline (\S\ref{subsec:nop-insertion}).
Finally, \sys %also\yifan{Remove ``also''?}
provides asynchronous decryption to accelerate
data transfer (\S\ref{ss:asyncdecrypt}).

%and two policies for ciphertext placement (\S\ref{subsec:design-allocation}).

\iffalse
%Moreover, \sys introduces two strategies to address inevitable prediction errors.
%In scenarios where a pre-encrypted \page is not required by the GPU,
%this \page is discarded and reverts to the GPU's standard online encryption process for confidential computing.
%The second scenario occurs when the sequence of pre-encrypted \pages is incorrectly predicted.
%Imagine a situation where the encrypted IV of a \page exceeds the current actual IV.
%In such cases, {\sys} inserts several smaller data \pages before the mispredicted \page.
%These \pages are then encrypted and sent to the GPU,
%incrementally raising the IV to match that of the incorrectly predicted \page.
%This approach ensures that the CPU resources spent on pre-encryption are not wasted.
\fi

We implement {\sys} with %\yifan{``by'' or ``with''?}
approximately 1K lines of code in C++. %the NVIDIA library and GPU driver. %\yifan{violates with impl}.
%Our performance experiments are carried out on an Intel server equipped with an NVIDIA H100-SXM GPU.
We conducted our performance experiments on an Intel server equipped with an NVIDIA H100-SXM GPU.
The evaluation results show that \sys significantly reduces the overhead associated with GPU \cc
for LLM serving and fine-tuning, cutting it from as much as 88.2\% to $<$19.6\% in throughput,
across various LLM sizes,
ranging from 13 billion to 175 billion parameters.

In summary, this paper makes the following contributions:
\begin{myitemize2}
  \item We conduct a %the first [cheng: tricky to claim this]
        comprehensive performance analysis of \nvcc on an H100 GPU enclave
         with LLM workloads.
  \item We propose {\techone}, an approach to greatly reduce the swapping overhead
  of \cc. It works well for LLM serving and fine-tuning.
  \item We have built a system {\sys} and evaluated its performance on multiple state-of-the-art
    LLM systems.
\end{myitemize2}

%- Challenge:
%  however, similar to CPU pipelining,
%  if the pre-encrypted data is incorrect (the GPU needs another piece of data)
%  then the entire pipeline has to be flushed and needs to start over.
%  - brief explain why like CPU
%    This is due to the encryption scheme.
%    To prevent replay attacks, each page is encrypted with an integer called IV;
%    IV needs to be agreed by the CPU and the GPU;
%    IV will increase by one for each encrypted page, which forms a total order of the pages.
%    Therefore, if a single incorrect page is encrypted, then all the following IVs in the pipeline
%    will be incorrect and have to be re-encrypted with the correct IVs.
%  - the challenge is to minimize the "pipeline flush", as it is super expensive in our case.

%- Observation:
%  LLM inferences are regular;
%  the page swappings are regular;
%  we can reliably predict which pages will be swapped in.

%- we introduce our system, \sys:
%  -- predict LLM needed pages
%  -- speculative pipelined swapping
%    -- model page swapping
%    -- data page swapping
%  -- fallbacks
%    -- XYZ
%  -- we update GPU driver with zero-modification
%    for applications or hardware.

%- contributions:
%  - study the bottleneck of confidential LLM inference
%  - propose speculative pipelined swapping, a new swapping technique
%    for near-zero confidential LLM inference
%  - build a system and experiment with it on multiple state-of-the-art
%    serving systems

\section{Background}
\label{sec:bg}

\subsection{Large Language Models (LLMs)}
\label{subsec:bg-llm}
\label{ss:llm}
% Goal of this subsection:
% 1. Introduce LLM. How LLM computes? => Part 1: Introduce transformer, self-attention and K/V Cache
% 2. Why swapping/offloading is an important technique? =>
%     Part 2.1: If model size could not fit in GPU
%     Part 2.2: If model size could fit in GPU, dynamic K/V Cache

% [cheng: cut the backgroun to 1page]
% * intro to LLM
% * intro LLM serving
% * intro to LLM training
% * KV cache
% * memory pressure
% * swappig

% Part 1
LLMs are language models that
take a sequence of text as input and produce a corresponding sequence of text as output.
The Transformer architecture~\cite{vaswani2017attention} is the predominant framework in LLM design,
used by well-known models like GPT~\cite{brown2020gpt}, OPT~\cite{zhang2022opt}, and LLaMA~\cite{touvron2023llama}.
An LLM typically consists of an input embedding layer, numerous transformer
layers, %(from tens to hundreds),
and an output embedding layer. 

% serving
The inference
process of LLMs begins with a prompt (a series of tokens).
Outputs are generated
iteratively, with each iteration processing the sequence through all layers to
produce the next token.
This iterative process, known as auto-regressive generation, continues until an
end-of-sequence token is produced or the output length reaches a threshold.
% [cheng: doesn't help the paper]
%Unlike traditional models like CNNs, which process only one iteration per request,
%LLMs feature auto-regressive generation, distinguishing them from these earlier models.

%% training
%%\cheng{TODO: briefly describe LLM training}
%LLM training is not auto-regressive but instead resembles traditional CNN
%training, involving one forward and one backward pass for each batch of input.
%However, the larger number of parameters in LLMs, compared to traditional CNNs, necessitates new training frameworks.
%While CNNs can be pre-trained on a single GPU over a few days,
%the significantly larger number of parameters in LLMs means that
%pre-training these models typically requires multiple GPUs and can extend the
%pre-training period to several weeks\cite{shoeybi2020megatronlm},
%making pre-training prohibitively expensive for many.
%As a result, fine-tuning LLMs becomes a more viable option because it requires far fewer resources and less time compared to pre-training.
%Research initiatives such as LoRA\cite{lora} focus on LLM-specific fine-tuning techniques.

Training an LLM from scratch is very expensive~\cite{shoeybi2020megatronlm}.
Instead, fine-tuning~\cite{lora} a pre-trained LLM is a more viable option for many
because fine-tuning requires far fewer resources and less time.
This paper aims at serving and fine-tuning LLMs with fast \cc.

\heading{\kvcache in LLM.}
\label{ss:kvcacheintro}
In LLM inference, the attention mechanism captures the relationship between the current
input token and preceding tokens by calculating their context as key and value
vectors. Storing these vectors in a \emph{\kvcache}~\cite{pope2022kv}
allows LLMs to save repetitive calculations,
significantly boosting inference efficiency. However, this
efficiency comes at the cost of increased GPU memory consumption.

% Part 2.1
\heading{Memory pressure in LLMs.}
The size of LLMs has been rapidly increasing, outpacing the growth in GPU
memory capacity. This disparity has made limited GPU memory a significant
bottleneck for serving and fine-tuning LLMs.
The memory pressure associated with LLMs originates from three primary sources.
First, both the inference and fine-tuning of LLMs require the entire set of
parameters for computation, and the model weights may exceed the GPU's
available memory capacity.
Second, for serving LLMs, \kvcache consumes a lot of memory,
for example, vLLM allocates about 30\% of GPU memory for \kvcache~\cite{vllm-paper}.
Third, activation and optimizer states during fine-tuning are memory-intensive.
The size of the activation state is proportional to the batch size, input length, and model hidden size,
while the optimizer state size depends on the number of trainable parameters.

%the dynamic properties of the {\kvcache} introduce additional complexity.
%In LLM serving systems, accurately predicting the memory requirement for the {\kvcache}
%is challenging due to the varying lengths of input prompts
%and the unpredictable output lengths that cannot be determined until the inference process is complete.

\heading{GPU memory swapping.}
To mitigate the LLM memory pressure,
previous studies have introduced various GPU swapping techniques.
FlexGen~\cite{flexgen-paper} only loads the parameters of the active layer onto the GPU,
while maintaining the parameters of other layers in CPU memory.
vLLM~\cite{vllm-paper} addresses memory shortages by pausing some running requests, swapping
their \kvcache to main memory, and later loading back the vectors to resume processing these requests.
%
%\cheng{TODO: what about training system's swapping?}
Deepspeed~\cite{deepspeed-zero} facilitates optimizer offloading and model
offloading to free up memory, enabling larger batch sizes and higher
throughput in pre-training and fine-tuning LLMs.

%In multi-GPU configurations, LLMs can distribute model weights and {\kvcache} across various GPUs,
%utilizing strategies such as tensor parallelism~\cite{shoeybi2020megatronlm}.

%Although our paper primarily focuses on single-GPU environments, the analyses
%and observations presented are also applicable to multi-GPU setups.

\subsection{Confidential Computing (CC)}
\label{subsec:bg-cvm}
% Goal of this subsection:
% 1. What new threat does cloud computing bring? => Part 1: Motivation of CVM
% 2. Does CVM introduce new challenge? => Part 2: CVM I/O

%Today's LLMs are commonly deployed on servers outfitted with high-end GPUs,
%an approach that entails significant financial costs.
%Cloud deployment, as an alternative, introduces a pay-as-you-go model,
%facilitating elastic GPU utilization and ensuring high availability for LLM services.
%Despite these advantages, concerns about privacy,
%such as the safeguarding of sensitive LLM model weights and the confidentiality of user data and responses,
%remain prevalent among LLM owners and end-users.
%Confidential computing emerges as a viable solution to these concerns.

Cloud computing is a fundamental component of today's digital world.
Confidential computing has emerged as a solution for securely outsourcing
computation to the cloud without compromising security and privacy.
Cloud providers like 
Microsoft Azure~\cite{azure-cc,azure-h100cc} and Google Cloud~\cite{googlecloud-cc}
have introduced \emph{confidential virtual machines (CVMs)}
as a key measure for confidential computing.
CVMs ensure robust isolation from untrusted hypervisors,
enhancing security and privacy with minimal changes to existing cloud infrastructure.

\heading{Confidential VMs (CVMs).}
A CVM, such as Intel TDX~\cite{tdxmodule} and AMD SEV-SNP~\cite{sevsnp},
isolates the user's OS (the guest) from the hypervisor (the host) and secures the user's
code and data without modifying applications.
Additionally, CVMs encrypt their memory to guarantee privacy.
CVMs generally incur minimal performance overhead,
averaging around 4\%~\cite{sevsnp-perf}.
I/O operations however may introduce significant performance overhead
due to data copying and encryption~\cite{li2023bifrost}.

\label{subsec:bg-h100}
\heading{Confidential Computing (CC) on GPUs.}
Beyond CPU-based CVMs, confidential computing on GPUs secures GPU computations
such as LLM serving and training.
Soter~\cite{atc2022soter}, designed for edge computing, uses
CPU-side confidential computing to eliminate the trust on GPU hardware.
It employs parameter morphing to offload certain layers to GPUs, ensuring model
weights secrecy. However, Soter imposes significant overhead (2$\times$ to 15$\times$)
on LLM inference tasks, as non-associative layers
like embedding must be processed on CPUs.
Alternatively, Honeycomb~\cite{osdi2023honeycomb} introduces a trusted software layer and
static validation to block insecure GPU commands, but this adds substantial
performance overhead due to extra runtime checks for indirect memory access,
heavily used in systems like vLLM~\cite{vllm-paper}.

% %In addition to CPU-side CVMs, the concept of confidential computing
% %extends to the GPU domain,
% Beyond CPU-side CVMs,
% confidential computing on GPUs
% offers protection for GPU computations like LLM serving and training.
% %
% Soter~\cite{atc2022soter} presents a system designed for edge computing scenarios
% that relies on CPU-side confidential computing,
% thereby eliminating the need to trust the GPU hardware.
% This system leverages a technique known as parameter morphing,
% which enables the offloading of layers with associativity properties to GPUs.
% This approach ensures that the plaintext of model weights remains secure and undisclosed to untrusted GPUs.
% But Soter has large overhead (2$\times$ to 15$\times$) on LLM inference tasks
% because non-associative layers, such as the embedding layers,
% has to be computed on CPUs.
% Honeycomb~\cite{osdi2023honeycomb} adds a trusted software component
% and uses static validation the block insecure commands to GPUs before submitting them.
% Honeycomb requires extra runtime checks for indirect memory access
% which is intensively used in LLM inference systems like vLLM~\cite{vllm-paper},
% bringing non-trivial performance overhead.
% % Example: The layer is y = k * x + b. The CPU sends 2 * k and 2 * b to the GPU,
% % And compute the layer on the GPU, and then divide the result by 2.
% % Drawback of Soter

% Goal of this subsection: Introduce H100 confidential computing
Unlike software-based solutions,
\nvcc relies on hardware:
\nv H100 GPU is the first commercial implementation with confidential computing
capability~\cite{h100cc-paper}.
Working with CVMs, H100 could build a GPU enclave, allowing users to have
exclusive control over the GPU and rejecting any access from the host, such as
read/write GPU memory and modify the control flow.
Hardware GPU \cc has low performance overhead
and is backward-compatible with existing applications.
This paper focuses on studying hardware GPU \cc.

\begin{figure}
  \includegraphics[scale=0.6]{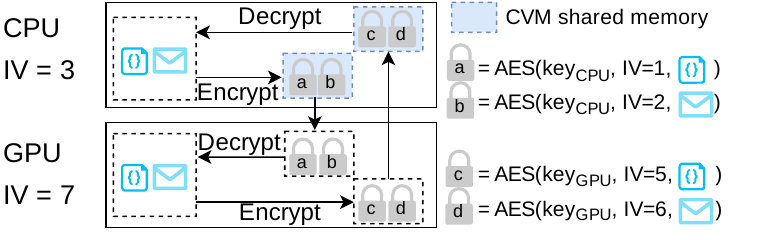}
  \caption{\textbf{Workflow of encrypted data transfer in NVIDIA CC.}
    The messages labeled ``a'' and ``b'' represent two consecutive ciphertexts
    transferred from the CPU to the GPU, while ``c'' and ``d''
    denote ciphertexts moved from the GPU back to the CPU.
    %\yifan{Should I add the following:
    After the transfers, the current IV of CPU and GPU is 3 and 7, respectively.
  }
  \label{fig:bg-h100cc}
\end{figure}

\heading{A closer look at NVIDIA CC.}
%Although NVIDIA \cc (CC) works with CVM,
%it does not use the same encryption key as CVM memory encryption~\cite{h100cc-paper}.
%Instead,
%[cheng: I won't expect they are using the same key...also, why starts with this information?]
%NVIDIA CC use different keys~\cite{h100cc-paper} 
%referred to as ``H100 key'', to encrypt data exiting the GPUs.
%
% Data must be encrypted using the H100 key before being transferred from the CPU to the GPU.
% Nvidia CC's encryption utilizes the AES-GCM algorithm and is similar to end-to-end encryption.
% The IV (also known as nonce) used in AES-GCM is incremented after each data transfer~\cite{h100cc-whitepaper}.
% The decryption side does not accept ciphertext generated by previously-used IVs, thereby preventing ciphertext reuse.
% % \TODO{Add a figure about how H100 CC works; why it is transparent}
%
Although CVMs encrypt their memory, this encryption is separate from that used
by NVIDIA CC. NVIDIA CC ensures the confidentiality
and integrity of communication between a CVM and a GPU via AES-GCM
encryption~\cite{h100cc-paper}.
A critical component of AES-GCM is the \emph{Initialization Vector} (IV),
%\yifan{Should we mention nonce here? Reviewer B asked about nonce: ``I am
%confused with the role of IV.'' Also, should we use \\emph here?}
a unique, non-repeating number (a nonce) required for each encryption session.
As we will show later (\S\ref{ss:iv}),
managing IVs presents a significant challenge.

Figure~\ref{fig:bg-h100cc} illustrates the workflow of data transfers of the NVIDIA CC.
Consider the process of copying memory from the CPU to the GPU;
the reverse process follows a similar pattern.
All CPU-side application data resides in the CVM's private encrypted memory.
To transfer data from the CPU to the GPU, the application invokes CUDA APIs
(e.g., \texttt{cudaMemcpyAsync}).
The CUDA library detects that it is operating in CC mode
and encrypts the data using the AES-GCM key and IV before sending it to the GPU.
The encrypted data is then copied to CVM shared memory, allowing the GPU to
perform DMA.

During data transfer, the IV is not explicitly sent.
Instead, the initial IV is synchronized during system
initialization, and both the CVM and GPU increment the IV after each transfer.
This design enables both sides to determine the current IV without additional synchronization.
Once the GPU receives the ciphertext, its copy engine---a piece of hardware---decrypts
the data and transfers the plaintext to GPU memory.
%\yifan{``copy engine'' is first mentioned here, without explanation. Should we add citation~\cite{h100cc-paper}
%or simply say dedicated hardware on GPU?}
The entire workflow is handled by the CUDA library and the GPU, requiring no
modifications to the application code.

\section{Analysis of LLMs on GPU Enclave}
\label{sec:analysis}
\label{s:bottleneck}

% nvcc on IO case
\nv reports that
compared with \cc disabled (\ccdisabled),
\cc enabled GPUs have significant performance
overhead on IO-intensive workloads like ResNet50 training~\cite{h100cc-paper}.
This is not surprising, as all I/Os need to be encrypted,
and the encryption is on the critical path.
On the contrary, LLM serving
%as well as training\CP{Yifan, is this right? we just said training has heavy overhead}
is supposed to be compute-intensive,
and hence has light overhead.
As shown by \nv, there is negligible overhead for serving BERT\cite{devlin2018bert} (a language model) on \ccenabled
GPUs~\cite{h100cc-paper}.

\begin{figure}[t]
  \subfloat[FlexGen]{
      \includegraphics[width=0.33\linewidth]{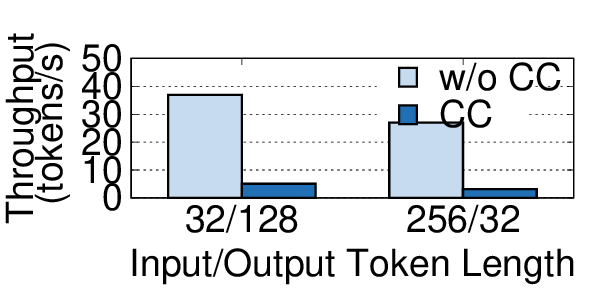}
      \label{fig:analysis-flexgen-opt-66b}
  }
  \subfloat[vLLM]{
      \includegraphics[width=0.33\linewidth]{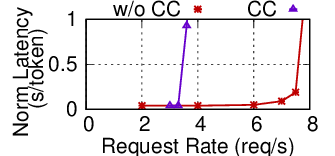}
      \label{fig:analysis-vllm-opt-30b}
  }
  \subfloat[PEFT]{
      \includegraphics[width=0.33\linewidth]{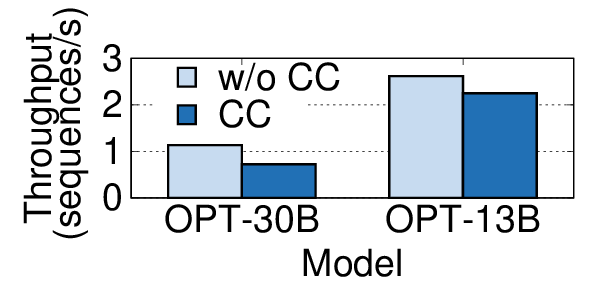}
      \label{fig:analysis-peft}
  }
  \caption{\textbf{\CF{\cc} overhead study.}}
\end{figure}

% nvcc on LLM serving
However, low-overhead serving or training is no longer true in LLM era.
Serving and fine-tuning LLMs is indeed compute-intensive, but it is even more of
a \emph{memory-intensive} workload.
We have observed that \ccenabled GPUs with 80GB GPU memory can have up to
an 88.2\% serving throughput drop on OPT-66B model (Figure \ref{fig:analysis-flexgen-opt-66b}),
a 36.2\% fine-tuning throughput drop on OPT-30B model (Figure \ref{fig:analysis-peft}),
and a 52.8\% serving capability drop on OPT-30B model (Figure~\ref{fig:analysis-vllm-opt-30b}).
This is due to the IO caused by memory swapping:
GPUs need to swap in and out memory for large LLM inference and fine-tuning.
We observe that there are two major reasons for swapping:
  (1) model offloading and (2) \kvcache swapping.
In this section, we dive into the details of the two reasons
with three state-of-the-art systems, \flexgen~\cite{flexgen-paper},
PEFT~\cite{peft} and
vLLM~\cite{vllm-paper}.

\heading{Microbenchmark.}
Before evaluating the performance of application workloads on \nvcc,
we conduct a microbenchmark to clarify the performance overhead of I/O operations.
This microbenchmark assesses latency and throughput across different I/O sizes. %including both small and large packets.[cheng: repetitive]
% The I/O tests are conducted by calling ``cudaMemcpyAsync'' API, which is non-blocking copy in non-CC environment.
We demonstrate memory copying from host to device, noting that the performance of the reverse operation is similar.

\begin{figure}[ht]
  \centering
  \footnotesize
  \begin{tabular}{@{}lr|cccc@{}}
  \toprule
  &   &   \multicolumn{4}{c}{I/O size} \\
  &   & 32B & 128KB & 1MB & 32MB \\
  \midrule
  \multirow{2}{*}{Latency ($\mu$s)} & \ccdisabled & 1.43 & 1.17 & 1.19 & 1.43 \\
  &\ccenabled & 14.93 & 22.809 & 162.5 & 5252.1 \\
  \midrule
  \multirow{2}{*}{Throughput (GB/s)} & \ccdisabled & -- & 27.16 & 48.2 & 55.31 \\
  &\ccenabled & -- & 3.32 & 5.82 & 5.83 \\
  \bottomrule
  \end{tabular}
  \caption{Host To Device Memory Copy of Different Data Size}
  \label{fig:analysis-microbenchmark}
  \vspace{-2ex}
\end{figure}

Figure~\ref{fig:analysis-microbenchmark} shows the result.
% In the table, "CC" represents the H100 with confidential computing disabled,
% and "CC" represents it with confidential computing enabled.
%        [cheng: this is redundant]
In the table,
``Latency'' measures the time from the invocation to the return of the host-to-device CUDA API,
while ``Throughput'' indicates the average throughput over 10K transfers.
%\XXX.\CP{Yifan, fill in; how long? like 5min?} \yifan{over 10,000 transfers}
We have omitted the throughputs for small data transfers (``32B'')
because the control-plane overhead is dominant in this case, and hence both throughputs are tiny.
% to a ``cudaDeviceSynchronize'' marking the completion of the I/O.
%
% The result shows the control-plane and data-plane overhead of \nvcc,
% indicated from API latency of small-sized data and throughput of large-sized data.
%        [cheng: communicate no infomation]
Notably, the throughput of a \ccenabled GPU is approximately an order of magnitude lower
than that of \ccdisabled, limited by the CPU's encryption capability.
Furthermore, the API latency for large packets remains relatively constant in
a \ccdisabled CPU but increases proportionally in a \ccenabled GPU.
This suggests that encryption and decryption processes are coupled with the API call.

\iffalse
From the result, we observe that:

\begin{myitemize2}
  \item For small packets, \nvcc incurs a significant overhead, likely due to metadata handling.
  \item For large packets, the throughput of \nvcc is approximately an order of magnitude lower than that of non-confidential computing, limited by the CPU's encryption throughput.
  \item TThe API latency for large packets remains relatively constant in non-CC environments but increases proportionally in \nvcc.
  This suggests that encryption and decryption processes are coupled with the API call.
\end{myitemize2}
\fi

% \TODO{Refine the figures}

Next, we examine three systems---\flexgen~\cite{flexgen-paper},
\vllm~\cite{vllm-paper}, and PEFT~\cite{peft}---to further
illustrate the bottleneck.

\heading{Case study 1: serving large LLMs with model offloading.}
% \TODO{Should I use training (fine-tune) in this case study?}
We first experiment with \flexgen and measure its throughputs.
\flexgen is the state-of-the-art LLM inference system that
supports serving large models whose sizes exceed the GPU memory capacity.
\flexgen prefers throughput over latency.
We run an OPT-66B model, which requires 132GB of memory,
exceeding the 80GB GPU memory capacity of the H100.
Thus, \flexgen needs to partially offload the model to CPU memory.
We use two configurations: input length 32 with output length 128,
and input length 256 with output length 32.
% One is to offload 60\% of the model weights with a batch size of 512.
% The other is to offload 50\% of the model weights with
% a batch size of 1024.
%\CP{Yifan, check the configs}
%\cheng{Yifan, fillin why you use the two configs; for example, are they what
%used by Flexgen?}
%\yifan{I have changed the config to show the best performance of input 512 and 1024,
%which is used in flexgen paper.}
% Both configurations select 32 as the output length.
% These are the configurations used in \flexgen's paper.
To study model offloading, we guarantee all \kvcache is stored in GPU memory to avoid \kvcache swapping.
Other configuration parameters are tuned to have the largest batch size for high throughput
on the \ccdisabled system.

Figure~\ref{fig:analysis-flexgen-opt-66b} shows the result.
\flexgen's throughput has dropped by up to 88.2\%
%\CP{Yifan, fill in a percentage here; the point is that we want to be consistent when talking about ``throughput drop''}
when \cc is enabled.
%(up to 91.9\% throughput drop
%compared the throughputs when \ccshort is disabled.
%
Note that \ccenabled and \ccdisabled GPUs have the same IO;
the difference is that the \ccdisabled GPU does not need to encrypt the data.
%
%We further study the time spent on each stage.
%\cheng{Yifan, measure the time spent on IO-transfer and encryption}.
%
Additionally, we measure that \flexgen uses about 64GB/s PCIe bandwidth when \ccshort
is disabled;
whereas with \ccshort enabled, it only uses approximately 6 GB/s bandwidth,
which is bottlenecked by encryption.

% \flexgen has already overlapped data transfers and computations.
% Larger batch size would increase the compute time per batch,
% and has significant performance improvement when \ccshort is disabled,
% but the performance on \cc only improves slightly.

%\cheng{for me, revisit the flow and move this paragraph.}
The above experiments show that
serving large LLMs that exceed available GPU memory
imposes significant overhead when \cc is enabled.
The primary reason is the
necessity for model offloading,
which in turn introduces significant IOs.
This, in turn, results in heavy encryptions,
ultimately leading to high overhead.

\heading{Case study 2: \kvcache swapping.}
Beyond large LLMs with model sizes exceeding GPU memory,
we observe that even when LLM models fit GPU memory, swapping still exists.
This is due to intermediate data management:
when LLMs are deployed (like serving as chatbots) and interact with the external world,
the dynamic inputs, outputs, and their intermediate data
trigger memory swapping, in particular \kvcache (\S\ref{ss:kvcacheintro}).

To investigate the mid-sized LLM serving,
we study \vllm, the state-of-art inference system
targeting low-latency LLM serving.
Unlike \flexgen,
\vllm does not offload model weights;
instead, it swaps intermediate data %including \kvcache
to handle the memory pressure
caused by the dynamics of serving multiple requests.
\vllm targets low latency.
We experiment \vllm with an OPT-30B model (60GB).
We configure \vllm with parallel sampling, a common decoding policy,
and ask \vllm to generate 6 output sequences for each input.

Figure~\ref{fig:analysis-vllm-opt-30b} shows the latency
changes when increasing request rates for \ccenabled and \ccdisabled GPUs.
When the request rate is low, they have similar performance
because there is no memory pressure, and memory swapping rarely happens.
This also implies that the encryption overhead for inputs and outputs is negligible.
However, as the request rate increases, the latency of the \ccenabled GPU
grows significantly.
This is due to the occurrence of intermediate data swapping.
%
%For serving 4 requests per second,
%\cheng{Yifan, fillin details: what's the latency? how much time spent on
%encryption and on data transfer? how much portion is due to the \kvcache?}
%
Taking a closer look,
we discover that the bottleneck primarily stems from memory ``swapping in'':
when the GPU requests data,
the CPU must encrypt the data before sending it to the GPU.
However, during the CPU encryption phase,
the GPU is idle due to the unavailability of the input.
These idle cycles affect the latency of the
ongoing request and other pending requests,
thus exacerbating the overall latency overhead.
%\cheng{Yifan, how long will the ``GPU idle'' be?
%put down data to support this claim}

From these experiments,
we summarize that when handling multiple requests,
serving even mid-sized LLMs with \ccenabled GPUs
imposes significant overhead.
The main cause of this overhead is the memory pressure caused
by the intermediate data (in particular, \kvcache) of multiple requests.
Furthermore, we note that the main bottleneck
is the CPU encryption when ``swapping in''.

% \cheng{Yifan, PEFT is just an technique. In our context, we need to name the
% system---a technique doesn't have overhead; a system does. I will just call it
% DeepSpeed-PEFT.
% %
% Also, why not just use LoRA instead of PEFT? What does PEFT brings us, but LoRA doesn't?
% DeepSeepd-LoRA will be easier to understand what's happening.}
% \yifan{PEFT is also a popular LLM fine-tune system: https://github.com/huggingface/peft.
% LoRA and DeepSpeed are indeed options of PEFT.
% LoRA is one fine-tuning approach, and DeepSpeed is one model offloading approach.}

\heading{Case study 3: fine-tuning LLMs.}
Fine-tuning is a widely used technique for customizing LLMs. In this
experiment, we aim to understand the performance bottlenecks of fine-tuning LLMs
in a \cc environment.
We experiment with Parameter-Efficient Fine-Tuning (PEFT)~\cite{peft}
with DeepSpeed~\cite{deepspeed-zero},
to evaluate the overhead brought by the \ccenabled environment.
Our experiments run LoRA~\cite{lora} on OPT-30B and OPT-13B models,
using the ultrachat dataset~\cite{ultrachat-dataset},
with the maximum batch size to trigger memory swaps.
% \CP{Yifan, what does this max batch size mean?}
% \yifan{Only model swapping is used in the experiments.
% Memory used by activation grows in proportion with batch size.
% Furthermore, PEFT does not support activation offloading.
% So too large batch size would make GPU memory overflow.
% So max batch size refers to the largest batch size that would not trigger GPU memory overflow.}

Figure~\ref{fig:analysis-peft} shows the result of PEFT.
The throughput drop brought by \cc is 36.2\% on OPT-30B model and 14.0\% on OPT-13B model.
The overhead is smaller on OPT-13B because it contains fewer parameters than OPT-30B
and has lower memory pressure, requiring less I/O.
Overall, the performance degradation stems from model offloading, similar to FlexGen.
% \CP{Yifan, check if this sentence is true} \yifan{Yes, this is true}

%\cheng{Yifan, why fine-tuning is put in-betwee two model serving?
%what's the problem of having FlexGen, vLLM, and then DeepSpeed?}
%\yifan{The flow is from model offloading to kvcache swapping.
%Within model offloading, FlexGen has higher overhead so it's placed to front.
%Maybe we can move kvcache swapping to front?}

%  - note: we would expect recompute \kvcache can help; the current vLLM
%  implementation doesn't have it.
% Note that vLLM proposes recompute as an alternative to K/V cache swapping,
% but recomputing has not been implemented on advanced decoding algorithms in vLLM.

%\section{Overview}
\section{Speculative Pipelined Encryption}
\label{sec:overview}

% [cheng: flow
%   - overall goal: improve perf, keep security, transparency
%   - what is ``security''?
%   - how to improve perf?
%     -- bottleneck is encryption
%     -- idea: move the encryption of the path
%     -- hard to do, why? IV
%     -- idea: predict IV regarding the patttern
%   - problem statement: predicting IV
%     -- observation
%     -- solution
%   - techon
%     -- challenges
%     -- high-level solution
% ]

%\heading{Goals.}
\label{subsec:ov-goal}
\label{ss:goal}

In this section, we describe our main idea,
the problem statement,
and technical challenges to build our system, \sys.
The primary goal of \sys is to minimize the overhead of \nvcc,
while preserving its (1) security guarantees and
(2) user transparency.
By user transparency, we mean that \sys applies to non-modified LLM
applications, including LLM models, deep learning frameworks, %serving \yifan{and fine-tuning} frameworks,
and any other supporting code and data. %\CP{Yifan, check if this is the definition of LLM app}
Next, we elaborate on \sys's threat model.

\heading{Threat model.}
\nvcc aims at protecting the confidentiality and integrity of
applications running on GPUs;
for LLM applications, these are
the model weights %[cheng: isn't this part of LLM app?]
and user request and response tokens.
{\sys} %adopts similar
shares the same %\CP{check; was ``similar''} 
threat model as \nvcc:
%and previous work on GPU confidential computing.
attackers can have full control over the hypervisor and the host OS,
but they cannot access the private memory protected by confidential VMs
and \ccenabled GPUs.
%
%they have physical accesses to over the host system.
%\TODO{Briefly introduce the thread model: CPU-side, GPU-side, CPU-GPU data transfer}
%
Like \cc on CPUs~\cite{tdxwhitepaper,sevsnp}, \nvcc is susceptible to
DOS attacks and side channel attacks;
\sys shares the same limitation.
In fact, \sys introduces new side channels, discussed in \S\ref{ss:sidechannel}.

\subsection{Main idea}

As mentioned in \S\ref{s:bottleneck},
the bottleneck when serving and fine-tuning confidential LLMs
is the encryption of memory swapping.
% in particular the ``swapping in'' process.
% \yifan{swapping in and out both has overhead.
% Swapping out (GPU-to-CPU) is overlapped using asynchronous decryption,
% which gains performance without the need of prediction.}
%
Our main idea is to somehow remove the encryption out of the critical path
of memory swapping.
Note that encryption is fundamental for security hence
cannot be avoided; it can only be hidden.
Hiding computation by pipelining is a well-established approach. %\CP{cites}.
%especially in GPU programming\CP{cites}.
%
We plan to pipeline encryptions, data transfers, and GPU computation for better performance.
%a canonical method in GPU programming. %\CP{cites}
%Figure~\ref{fig:overview-mainidea}
Figure below
%\yifan{Could we use unlabelled figure?} 
%[cheng: yes, that's fine; this part is very integral to the paragraph]
visualizes our main idea along with a comparison to \nvcc.
On the left is an example of toy code; on the right is an illustration of how \nvcc and \sys execute it.
%The comparison highlights the data swapped between the host and device.

%\begin{wrapfigure}{r}{0.4\textwidth}
\begin{figure}[h]
    \includegraphics[scale=0.6]{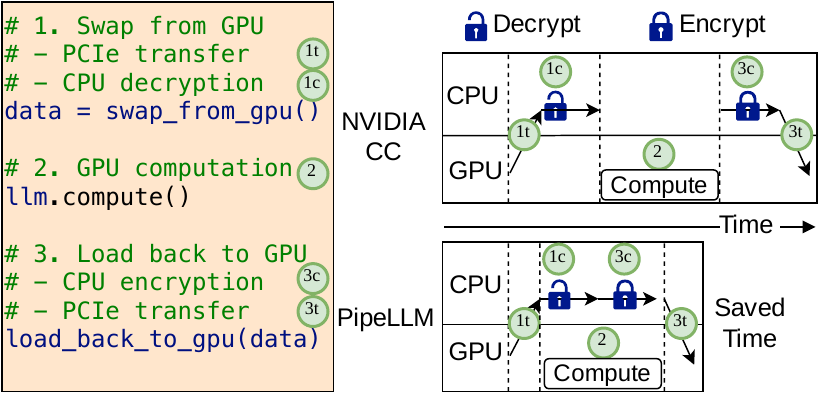}
\end{figure}
%\end{wrapfigure}

For transparency, \nvcc performs on-the-fly encryption
and decryption (indicated by ``1c'' and ``3c'') within its interface (e.g.,
\texttt{cudaMemcpyAsync}), together with data transfer steps (``1t'' and ``3t'').
\sys decouples and pipelines data transfer, encryption, and decryption to
minimize GPU idle time.
Note that GPU-side encryption and decryption are not depicted.

A major challenge to realize \sys's pipelining
is pre-encrypting data (i.e., the ``3c'').
\sys must ``predict''
which data the GPU will request
and, crucially,
the precise order of these requests.
This sequence is essential due to cryptographic requirements,
which we elaborate on below.

\label{ss:iv}
\heading{Swapping sequence and IVs.}
%\yifan{Merge with background} [cheng: no, we're good]
\nvcc (H100) encrypts I/Os using AES-GCM, %AES256-GCM, [cheng: be consistent]
a stream cipher that requires an \emph{Initialization Vector (IV)}
which is an integer.
In this encryption scheme,
each encryption needs three pieces of inputs:
(a) a piece of plaintext,
(b) an encryption key, and 
(c) a unique IV;
the output is the ciphertext.
It is crucial that
the IV for each encryption differs from the previous ones.
% and it remains a secret between the encryptor and the decryptor.
% Reusing IVs across multiple encryptions poses a risk of leaking the key.
%
In the H100 implementation, the CPU and the GPU agree on a starting IV;
for every following encryption, the IV is increased by 1
(IV is recorded in cyclic code in \nvcc, but we use decimal numbers to describe IV in this paper for comprehensiveness).
To pipeline encryption and data transfer,
one needs to somehow guess the sequence of memory swapping
a priori and pair the sequence with the increasing IVs.
Next, we define the memory-copy prediction problem.

% \heading{H100 technicality.}
% H100 confidential computing uses AES256-GCM for end-to-end encrypted transfer.
% AES256-GCM is a stream cipher, which requires a 96-bit IV (Initial Vector) to support the encryption process.
% The IV of each encryption should be kept different, or the key would be leaked.
% In H100, GPU-side IV is configured to increment 1 per transfer.
% NVIDIA does not provide a command for CPU-side driver to manually configure GPU-side IV.
% NVIDIA points out that the CPU-side encryption would bring as much as 45\%
% performance overhead in ResNet50 training cases
% compared with the case without confidential computing.
% Meanwhile, the GPU-side encryption would not bring overhead because the GPU uses tailored hardware for encryption
% which could provide encryption throughput as PCIe line rate.
% This work will analyze and optimize the performance of NVIDIA confidential computing in LLM inference use cases.

%\subsection{Definition of Prediction Success and Failure}
\subsection{Problem Statement}

% The user asynchronously submits a set of memory copy commands:
% \(\{t_1, t_2, ..., t_n\}\).
% We can assume that the commands do not overlap because of the asynchronous property.
% The predictor prepares an ordered sequence of ciphertext: \((p_1, p_2, ..., p_m)\).
% \(t\) and \(p\) refers to truth and prediction respectively.

% We define the success as:
% \[m \ge n\ and\ \{t_1, t_2, ..., t_n\} = \{p_1, p_2, ..., p_n\}\]

% Note that the order of \((p_1, p_2, ..., p_n)\) does not make sense because any two commands do not overlap.

% \noindent\textbf{F1: data loss.}
% \[\exists i \in [1, n], t_i \notin (p_1, p_2, ..., p_m) \]
% The preditor fails to predict the swapped data \(t_i\).

% \noindent\textbf{F2: wrong sequence.}
% The prediction does not meet the condition of prediction success, and \[\{t_1, t_2, ..., t_n\} \subseteq \{p_1, p_2, ..., p_m\}\]

For an LLM application,
when swapping memory (e.g., model weights and \kvcache),
the LLM system submits a sequence of memory-copy commands (\texttt{cudaMemcpyAsync});
it then uses a synchronization API (\texttt{cudaDeviceSynchronize})
to ensure the data has been loaded into the GPU's memory;
finally, it starts computation on the GPU.
Note that other than memory swapping, there are
memory-copies between CPUs and GPUs for normal computations.

The problem to solve is to predict the
next several memory copies and their IVs,
given the current IV
and the execution context.
%from a GPU driver's perspective.
%
Since there is no modification to the LLM applications,
only low-level memory-copy information is available.
We assume LLM models %and the serving systems (e.g., \flexgen and \vllm)
are known.

Predicting GPU memory copies in general is hard.
For serving and fine-tuning LLMs, we want to ignore normal memory copies
and \emph{only} predict memory swaps
 % \TODO{memory swappings?}
 % \cheng{Memory swap can be a countable noun; swapping describes the procedure/process, which is usually not countable.}
because (i) memory swaps contribute to the most of encrypted data transfers
and are the performance bottleneck (\S\ref{s:bottleneck});
(ii) the swapped memory remains unchanged on CPU, hence can be encrypted
ahead of time without worrying the contents will be modified;
(iii) LLM inference and fine-tuning are highly regular (\S\ref{ss:llm});
thus, accurately predicting memory swapping sequence is possible.

\label{ss:observation}
\label{ss:observations}
\heading{Observations.}
To predict memory swaps,
we observe that the sizes of LLM memory copies have the following patterns:
(1) the size of memory swapping (usually $>$128KB) is significantly larger
than other data transfers (usually $<$8KB).
(2) we can distinguish model offloading and KV cache swapping by calculating their sizes ahead of time based on the target LLMs.
%\yifan{Change to ``working'' or ``target''?} LLMs.
% (2) we can distinguish model offloading and \kvcache swapping
% by identifying parameters during application initialization (because parameters are allocated on or loaded into GPU before training or inference start)
% and treating other swapped data as \kvcache.
The two patterns allow us to accurately classify memory copies.
%If a mis-classification happens, it would be fixed in error-handling be recorded by \sys for higher accuracy in later classifications.
%
%the size of each model layer and each \kvcache
%block can both be calculated
%from the model structure, so they have strong relation.
%
In addition,
most of the transferred data is memory swaps
which are read-only.
This enables an efficient validation by leveraging page faults
(we describe details below, \S\ref{ss:s5-validation}).
%
%CPU does not modify the swapped data,
%which can be used for determining if a memory is indeed a swap.
%
Finally, we observe that
performing a ``NOP'' data transfer (a 1-byte dummy)
increments the current IV, with low overhead.
This implies that predicting a larger IV than the ground truth
has little penalty.

\subsection{Speculative Pipelined Encryption}
\label{ss:techone}

% [cheng: we have to give a crisp description of the prediction b/s readers are
% keen to see what we're doing]
At a high level,
\sys's prediction policy repeats the swapping patterns it has seen before,
and predicts swaps with slightly larger IVs as a leeway
for compensating other data transfers.
For example, \flexgen almost always swaps transformer layers in order.
So, when transferring layer $i$, \sys predicts and encrypts layer $i+1$.
As another example, \vllm uses FIFO (First-In, First-Out)
and LIFO (Least-In, First-Out) to swap \kvcache for different chunks.
\sys will predict swaps by FIFO and LIFO accordingly.

With the above prediction,
\sys introduces \emph{\techone} to
\emph{predict} a sequence of swapped memory chucks in the near future
and
\emph{pipeline} the encryptions and ciphertext transfers
to hide the bottleneck.
Furthermore, \sys develops a set of techniques to lower the penalty
of incorrect predictions.
The \techone can be divided into three stages:
\begin{myenumerate2}

    \item \emph{Prediction stage}: in this stage, \sys detects which GPU memory
        belongs to model weights and \kvcache without looking into the
        application's code.
        It further predicts the sequence of memory chunks to be swapped,
        and encrypts them with corresponding IVs.

    \item \emph{Validation stage}: at the moment of the application submitting
        the memory-copy commands,
        \sys validates the pre-encrypted data
        by checking if the data is correctly encrypted
        (e.g., no update to the plaintext)
        and if using the right IVs.
        If the check fails, \sys goes to the error-handling stage.

    \item \emph{Error-handling stage}:
        when the requested sequence is different from the predicted sequence,
        \sys needs to relinquish the current pipeline
        and start with the ground-truth sequence.
        %\sys applies multiple optimizations to this stage (\S\XXX).

\end{myenumerate2}

%\cheng{Later, term check: failures vs. errors}

\heading{Technical challenges.}
\sys faces multiple technical challenges.
We list them below and address them in the next section (\S\ref{sec:design}).
First, \sys has no knowledge about high-level semantics
(e.g., which piece of data to swap), %\CP{for me, rethink if this is challenge or the problem itself?}
%
%\sys's performance highly relies on prediction accuracy.
%One possible approach is to provide extra API for
%application to guide the prediction.
%
%High-level semantics will help predictions.
which significantly helps predictions.
However, getting them requires modification to the application and violates the
user-transparency goal (\S\ref{ss:goal}).
Therefore, {\sys} has to make predictions based on limited low-level information,
including LLM features and API trace of applications.

Second, \sys needs a computational and memory efficient validation.
Validation checks whether the pre-encrypted data aligns with what is requested.
The challenge is how to efficiently check if the data has been updated.
Again, note that \sys has no information about whether the data is read-only
(e.g., swapped data) or read-write (e.g., data requiring CPU updates).
One straightforward solution is to store
(a) the plaintext (that might be updated),
(b) the pre-encrypted ciphertext, and
(c) the original plaintext associated with the pre-encrypted ciphertext.
By comparing (a) and (c), \sys will know if (b) is valid.
It however requires data comparison on the critical path,
which increases runtime overhead.
This solution also triples the memory usage.

\iffalse
\sys should not use much more CPU memory
compared to the baseline, \nvcc.
This is challenging because
during the validation stage, \sys needs to compare
the predictions with the requested data.
A straightforward implementation is to
store both the plaintext and ciphertext of all data,
and to compare the predicted plaintext and data
from copy command.
\cheng{I don't understand this part.
Doesn't the GPU use an address or some identifier to fetch the data?
what is ``data from copy command''?}
\yifan{the copy command involves a CPU base address and size.
The data is [base, base + size).
Python programmers may write ``x.copy\_to\_gpu()'', and are not aware of the address.
I want to fill in the gap between ``a bunch of data'' and ``address, size''.}
\fi

%At the validation stage, \sys checks the correctness of the predictions,
%involving that the data is encrypted.
%A challenge arise from the validation of whether the data has been predicted
%and encrypted by \sys.
%One trivial solution is to let \sys store both the plaintext and ciphertext,
%and to compare the predicted plaintext and data from copy command.
%However, it adds another data comparison and increases runtime overhead on CPU.
%Moreover, offloading technology in LLM consumes lots of memory, so {\sys}
%should limit its memory usage.
%This solution adds triples the memory usage to for prediction.
%{\sys} should limit its memory usage.

Third, \sys needs to handle prediction errors efficiently.
Pipeline relinquishing is expensive.
On the one hand, when encountering an error,
the wanted data needs to be encrypted on-the-fly,
leading to high tail latency.
On the other hand, the encrypted data remaining in the pipeline
cannot be used due to incorrect IVs,
which requires another round of encryption for all the subsequent data.

%The prediction system could not ensure 100\% accuracy.
%Even with high accuracy, {\sys} should have a mechanism for prediction failure.
%On one hand, the failure handling should be fast to avoid high tail latency.
%On the other hand, attackers may potentially obtain information from the
%failure data and the transferred data,
%so it should avoid leaking sensitive information to attackers.
%  [cheng; security part is not a "challenge"; it is simply "incorrect"]

%[cheng: why is this a challenge? isn't this the main problem, namely, how to
%  accurately predict?]
%Finally, Dynamicity of small size data transfer.
%Small size data transfer brings nearly no overhead to \cc, but it would greatly
%affect the predition. The reason is, each data transfer would increment the IV.
%Take \kvcache swapping as an example. A swapped \kvcache vector would not swap
%in until other running requests finishes or swaps, freeing enough space for it,
%which would happen in several iterations. However, small size data transfer
%happens in every iteration of LLM inference, which makes it very hard to
%predict the IV of the swapped \kvcache.

\section{\sys Design}
\label{sec:design}

\newcommand{\prediction}{prediction\xspace}
\newcommand{\predict}{predict\xspace}
\newcommand{\encworker}{encryption worker\xspace}
\newcommand{\allocation}{ciphertext placement\xspace}
\newcommand{\bounce}{bounce ciphertext placement\xspace}
\newcommand{\zerocopy}{zero-copy ciphertext placement\xspace}
\newcommand{\inplace}{inplace ciphertext placement\xspace}

\begin{figure}[t]
    \includegraphics[width=\linewidth]{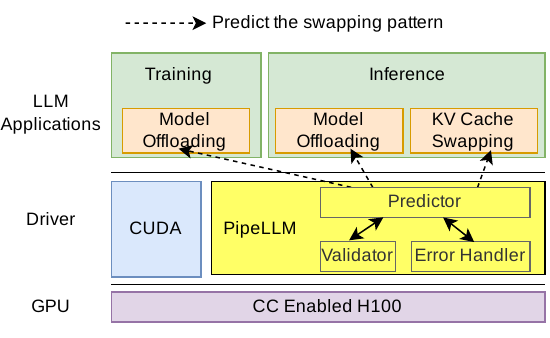}
    \caption{\textbf{\sys's architecture.} Yellow parts are \sys.
    % \cheng{(1) Be consistent: either pre-encryption (Fig4) or speculative encryption
    % (2) in fact, I found ``pre-encryption'' and ``predictor'' have a lot to do with each other
    % (shall they have some arrows or something?)
    % (3) based on the description below, predictor, validator, and error handler
    % should be three different pieces. Better to separate them.
    % (4) It might be easier to have a box called ``\sys'' with different component boxes internal.
    % }
    % \yifan{
    %     (1) ``pre-encryption'' is removed from Fig4.
    %     Another issue: I removed async decryption from the figure,
    %     because the paragraph ``\sys architecture'' did not mention it.
    % }
    }
    \label{fig:overview-arch}
    %\vspace{-2ex}
\end{figure}

\heading{\sys architecture.}
We show the architecture of {\sys} in
Figure \ref{fig:overview-arch}.
{\sys} works on the library and driver 
level, %\CP{Yifan, check: we're not doing driver, right?}
requiring no modification to user
applications.
{\sys} consists of a predictor (\S\ref{ss:predictor}),
a validator (\S\ref{ss:validator}), and an error handler (\S\ref{ss:errorhandler}).
The predictor produces the next sequence of memory chunks to swap,
based on the low-level information observed by \sys;
the validator checks if the pre-encrypted data is valid before sending it to the GPU;
the error handler re-plans the pipeline when facing a prediction error.
%\sys uses its own memory allocator to minimize the memory usage.
%
%Encrption window and inplace encryption is proposed based on \textbf{O1.1} and
%\textbf{O1.2} respectively.
%
%Next, we elaborate on
%the main components of \sys:
%predictor (\S\ref{ss:predictor}),
%validator (\S\ref{ss:validator}),
%error handler (\S\ref{ss:errorhandler}),
%and asynchronous decryptor (\S\ref{ss:asyncdecrypt}).

\subsection{\sys Predictor}
\label{subsec:design-prediction}
\label{ss:predictor}

% - prediction formalization
% - detailed prediction
% - other ios

\sys's predictor predicts which memory chucks to be swapped in and the chunks' corresponding IVs.
In practice, the swapping requests are in batch:
a set of memory copy operations (i.e., \texttt{cudaMemcpyAsync}) followed by a
synchronization (i.e., \texttt{cudaDeviceSynchronize}).
The LLM systems such as \flexgen, \vllm, and \peft need to maintain the order between
batches, without enforcing a %\yifan{the?}
specific order within a batch.

The predictor takes in the following three inputs:
(1) a swapped-in batch history $[B_0,\cdots,B_n]$,
in which each $B_i$ contains a set of memory chucks $\{C_p, \cdots, C_q\}$
and $B_n$ is the most recent swapped-in batch;
(2) the swapped out memory blocks $\{C_i, \cdots, C_j\}$ on main memory
which haven't been swapped into the GPU yet,
and (3) the current Initialization Vector (IV), $IV_{cur}$.
% \yifan{Should we add \\emph here?} [cheng: emph only need to be used when the term appears for the first time]
The predictor can be abstracted as a function $f$
that predicts the next swapped-in block $C_{next}$ and its corresponding $IV_{next}$:
\[
    f([B_0,\cdots,B_n], \{C_i,\cdots,C_j\}, IV_{cur}) \to (C_{next}, IV_{next})
\]
So far, there is a limited number of swapping patterns in today's systems,
including repetitive pattern (e.g., model weights offloading),
FIFO (e.g., layer-wise \kvcache swapping), and LIFO (e.g., request-wise \kvcache swapping).
We elaborate on these patterns below.
However, \sys's predictor is general and can easily extend to other patterns.
To implement a new pattern, one needs to recognize the pattern from the history
and write a prediction function given the current swapping states.
Our future work is to use ML models to learn the predictor $f$ without human efforts.

%\heading{IV \prediction.}
% Model weight offloading in LLM training or inference systems typically follows a consistent pattern,
% making the offloading trace identical among iterations and predictable for \sys.
% The model can either be fully offloaded or partially offloaded (where part of the model is offloaded).
% In the case of full offloading, the process adheres to the layer-by-layer structure of model execution.
% In partial offloading, LLM applications typically offload the same set of parameters across iterations
% given that each transformer layer has identical structure and size
% For example, if an application offloads parameters from layers 1 and 3 in one iteration,
% it will likely continue to offload these specific layers in subsequent iterations.
% This consistency makes the offload pattern highly predictable for \sys.
% Therefore, for model offloading,
% \sys tracks the offloading history of all offloaded blocks.
% Later, \sys predicts an IV and pre-encrypts the upcoming blocks of parameters.
% \sys learns whether the model is in forward or backward pass, adapting to the sequential and reverse sequential patterns, respectively.

\begin{figure}[t]
    \subfloat[Repetitive pattern] {
    %\subfloat[Layer-wise prediction] {
        \includegraphics[width=0.5\linewidth]{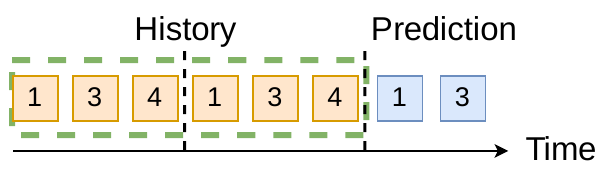}
        \label{fig:design-predict-layer}
    }
    \subfloat[LIFO (last-in-first-out)] {
    %\subfloat[Request-wise prediction] {
        \includegraphics[width=0.5\linewidth]{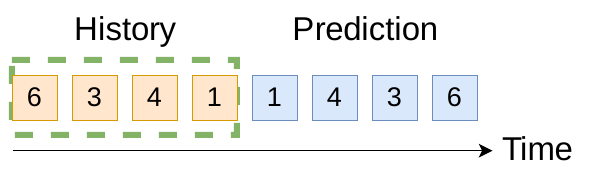}
        \label{fig:design-predict-request}
    }
    \vspace{1ex}
    %\caption{\textbf{Prediction of \sys}}
    \caption{\textbf{Common patterns in LLM memory swapping.}}
    % \textmd{X-axis is configuration of FlexGen.}
    \label{fig:design-predict}
    %\vspace{-4ex}
\end{figure}

\heading{Current swapping patterns.}
In the following, we describe several swapping patterns we observed in today's LLM systems.
For model offloading like \flexgen,
it has a repetitive pattern.
Figure~\ref{fig:design-predict-layer} gives an example.
Each box represents a layer, tagged by the layer number in the model (e.g, the first layer is tagged by 1).
Parameters from layer 1 are first reloaded,
followed by layers 3 and 4, and then back to layer 1.
This pattern indicates that layers 1, 3, and 4 are offloaded.
%Other layers, like Layer 2, do not appear in the history
%because they are computed immediately after the immediate previous layer,
%therefore are not offloaded.\CP{I don't understand this sentence}
%
Given that the most recent layer in the swap history is layer 1
(the final red block in the Figure~\ref{fig:design-predict-layer}),
\sys predicts that layer 3 will be the next to be reloaded and prepares it with an appropriate IV for encryption.
We observe that model offloading in most applications typically follows this
repetitive pattern.
Since each layer in LLMs shares the same structure and number of parameters,
it is unlikely that applications offload a particular layer in some
iterations while retaining it in the GPU during other iterations.

Another major swapping in LLMs is \kvcache swapping,
typically conducted on a layer-wise or request-wise basis.
%both of which are highly predictable patterns.
In layer-wise swapping, the application decides whether to swap the \kvcache
generated in each layer for all requests in the batch.
Alternatively, request-wise approach involves swapping all layers of \kvcache
from certain requests in case of memory shortage to free up memory for others.
Layer-wise swapping aims to achieve high throughput as it allows the batch size to remain unchanged;
request-wise swapping reduces latency for high-priority requests and lowers the average latency overall.

%Layer-wised swapping follows a consistent pattern and is highly predictable for \sys.
% For \kvcache swapping, the swapping policies
% of existing systems---for example, Flexgen~\cite{flexgen-paper},
% vLLM~\cite{vllm-paper} and DeepSpeed~\cite{deepspeed-paper}---are
% either FIFO or LIFO.
% This makes sense, as the swapped \kvcache is either layer-wised---swapping
% \kvcache of each layer in order, hence FIFO---or
% associates with requests:
% swapping out all \kvcache for one user request.
% Request-based swapping is likely to be LIFO 
% because the first swapped-out request has the lowest priority, so it is swapped in last.
% \sys learns the current swapping pattern and predicts near-future swaps by
% following the pattern.
% For example, if the policy is FIFO and memory chucks are
% swapped out in the order of \texttt{chunk A}, \texttt{chunk B}, and \texttt{chunk C};
% \sys then pre-encrypt in the order of \texttt{chunk A, B, C} with corresponding IVs.

In layer-wise swapping, applications swap out \kvcache of each layer in order,
and then retrieve them in the same order,
thus the pattern is FIFO.
In request-wise swapping, applications usually swap out the lowest
priority request in the current batch when facing memory shortage.
Therefore, the first request to be swapped out is likely the one with the
lowest priority among the requests, and it will be the last to be
reloaded. So the swapped in pattern is LIFO, as depicted in Figure~\ref{fig:design-predict-request}.

%\sys can learn and predict this pattern with a high degree of accuracy.
%Note that \sys's approach applies to other regular \kvcache swapping policies,
%and can be easily extended to future systems.

\label{ss:smallio}
\heading{Targeting major memory swapping.}
The predictor focuses on major GPU memory swaps---model weights offloading and
\kvcache swapping.
Other than the two, I/Os %DMAs
also involve other small data transfers,
for example, user input and output tokens.
As mentioned in \S\ref{sec:analysis},
since the encryption of them bring negligible overheads,
\sys does not pipeline them.
% \sys predicts data swapping based on the swapping history.
% The size of model layer and K/V cache have some relation and feature,
% so whether a request is model weights or K/V cache can be implyed based on the size.
By the model definition (i.e., its computational graph and weights),
\sys distinguishes the data of model weights and KV cache with high accuracy
without additional hints from LLM applications.
% \sys identifies parameters when they are allocated on or loaded into GPU.
Therefore, \sys can distinguish the data of model weights and \kvcache with
high accuracy without additional hints from LLM applications.
Therefore, \sys knows if a memory swap is model offloading
or \kvcache swapping with high confidence.

Notably, those small data transfers would make the IV increment.
Since swapping and small I/O are intertwined and interleaved in applications,
\sys would predict a larger IV for swapping to handle IV increments by small I/O.
If predicted IV is larger, the ciphertext could still be utilized for transfer,
which would be explained in \S\ref{ss:errorhandler}.

\subsection{Validator: Validating the Consistency of the Ciphertext}
\label{ss:s5-validation}
\label{ss:validator}

\heading{The problem.}
As mentioned,
\sys speculatively encrypts memory chunks that haven't been requested by the GPU.
The problem however is, how does \sys know if the original
data has been updated when later requested?
For example, consider the case that
an application transfers some data from the GPU to the main memory,
updates the memory in-place,
and then writes it back to the GPU.
In this case, this method could result in sending ciphertext of
outdated data when speculative encryption is performed before data update by the application.
Therefore,
before sending ciphertext,
\sys must verify that the corresponding plaintext matches the expected data.

\heading{Our method.}
\sys's validator uses
an efficient
% \(O(1)\) [cheng: this is not about complexity]
approach to validate the encrypted data.
The validation works as follows.
During the prediction stage, \sys labels each pre-encrypted ciphertext with
the corresponding plaintext's address range on the CPU memory.
\sys then revokes the write permission of these memory pages.
Therefore, when facing data modifications by the application,
\sys detects the changes by a page fault.
The page protection can thus prevent \sys from violating the system's correctness.
During the validation stage at runtime, \sys checks if the address and length
of the swap align with the associated labels to determine the match between plaintext
and ciphertext.
This method is highly effective for LLMs
because applications do not move or update the swapped model weights or \kvcache in the CPU memory.
The pre-encrypted ciphertext only involves large memory transfers which are typically
non-modified swaps.
Therefore, page faults rarely
%\yifan{change to ``rarely''?}
occur and thus bring little performance overhead.

% \begin{figure}
%     \includegraphics[scale=0.5]{./figures/drawio/cipher-validation.pdf}
%     \caption{Ciphertext validation.}
%     \textmd{\sys only records the base addresses and sizes of swapped data blocks.
%     The validation checks the two, consuming \(O(1)\) time.
%     Application's write to predicted swapped blocks means the prediction is wrong.}
%     \label{fig:design-cipher-validation}
% \end{figure}
% % \TODO{A figure to describe ciphertext validation}

% It is implemented by MPK/PKU for thread-level protection.

\subsection{Error Handler: Tolerating Wrong IVs with Re-Ordering and NOP Padding}
\label{subsec:nop-insertion}
\label{ss:errorhandler}

% Re-order example:
% Prediction block number: 1 2
% Request block number: 2 1
%% Request block 2 => prediction larger than current, pending
%% Then, request block 1 => prediction ok

\sys may incorrectly predict the swapped memory chunks, their sequence,
or their IVs.
Some errors are ``irrecoverable'': for example, if the predicted IV is smaller
than the current IV, the pre-encrypted data must be discarded.
In such cases, \sys's error handler resets the pipeline by discarding the speculative
encrypted memory and restarting the process.
However, many errors are recoverable and can be corrected with specific modifications
to maintain pipeline integrity.
We introduce two approaches---swap re-ordering and NOP
padding---to address these errors.
We experiment with various incorrect predictions in \S~\ref{ss:wrongpredicrion}.

\heading{Swap re-ordering.}
Swap re-ordering helps the case that
\sys incorrectly predicts the order of memory blocks within one batch.
% LLM applications reload some data blocks at a certain IV,
% but \sys predicts at a IV higher than current.
% % The opposite case would not happen because \sys drops predictions whose IVs
% % are smaller than current for old IVs are never used in the future.
%
As mentioned in \S~\ref{ss:predictor},
LLM applications may batch multiple swap-in memory blocks,
in which a synchronization (e.g., \texttt{cudadevicesynchronize})
marks the boundary of each batch.
\sys can re-order these speculative encrypted blocks within the batch,
without having to relinquish the pipeline.

%would first pend a swapping request with wrong IV,
%and re-order disjoint swapping requests to match the predicted IV.
%Thanks to the efficient validation time, the pending brings little overhead.

% NOP example:
% Prediction block number: 1 2
% Request block number: 2
% \sys: NOP => swap 2 and drop 1 (actually, predict 1 at a higher IV)

\begin{figure}
    \includegraphics[scale=0.45]{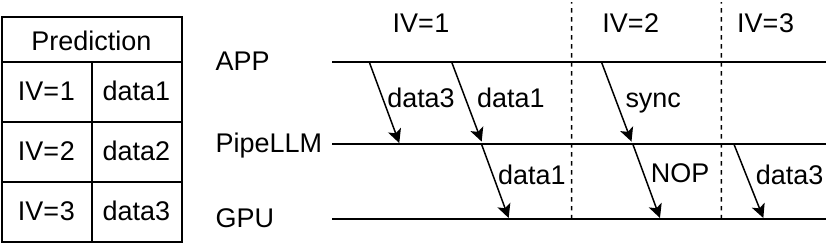}
    \caption{\textbf{Tolerate wrong IVs with re-ordering requests and padding NOPs.}
    In this figure, the current IV starts with 1, and the application
    requests ``data3'' which is speculatively encrypted by IV=3.
    Since IV=3 is greater than the current IV=1,
    \sys suspends this request. %until a synchronization (``Sync'').
    %The system's IV would grow with successful predictions.
    The next request is ``data1'' which is encrypted by IV=1, so \sys sends ``data1'' and IV advances to 2.
    Finally, by synchronization ``sync'', \sys sends a NOP to increase the current IV=3,
    invalidates the incorrectly encrypted memory ``data2'',
    and commits the pending requests ``data1'' with IV=3.
    }
    % \TODO{Polish this figure}}
    \label{fig:design-tolerate-wrong-iv}
    %\vspace{-3ex}
\end{figure}

\heading{Padding NOPs.}
%After re-ordering, there might still be some pending requests, which is solved by padding NOPs in the design of \sys.
%A 1-byte data could be send to increase the current IV with small overhead,
%and \sys could ultilize the predicted ciphertext without re-encryption even if the predicted IV is larger.
%
A NOP is an operation to send a 1-byte piece of data to increase the current IV.
Padding NOPs is useful when
\sys encrypts a data block with an IV that is larger than it should be.
For example, in Figure~\ref{fig:design-tolerate-wrong-iv},
``data3'' is speculatively encrypted using IV=3;
however, current IV=2.
\sys can forward IV by sending a NOP.
In addition, NOP padding enables \sys to tolerate small-sized data
transfers (mentioned in \S\ref{ss:smallio}) without restarting the pipeline.
%compromising the confidentiality and integrity of the \cc.

% \TODO{Discussion: Re-ordering does not bring overhead.
% 1 NOP costs 10us, while reencrypting 1MB costs 232us.
% \sys can combine those two methods.}

\subsection{Asynchronous Decryption}
\label{ss:asyncdecrypt}

So far, our discussion has primarily focused on CPU-to-GPU data transfers and
CPU-side encryption. Regarding CPU-side decryption during GPU-to-CPU data
transfers, we observe that by default, decryption is unnecessarily synchronous.
In particular, LLM applications do not alter swapped-out model weights or \kvcache.
%except during the swapping process itself.
%Additionally, some data may be
%processed on the CPU after being retrieved from the GPU.\CP{check with Yifan for the two sentences}
Based on this observation, \sys implements asynchronous decryption for swapped-out data.
The memory copy returns before decryption and leaves the data at
the destination address unmodified.

As a technicality,
the async decryption works as follows.
\sys determines whether the data being transferred is part of the \kvcache or model parameters
based on their feature on data size mentioned earlier.
If yes, \sys decrypts it asynchronously for better overlapping;
otherwise, \sys does the default, decrypting the data immediately.
%based on our observation of classifying swapped data (\S\ref{ss:observation}).
%%% In training, GPU might modify the weights
% It's important to note that the GPU
% does not modify the model weights; therefore, any swapped model weights
% are simply discarded rather than copied back to the CPU memory.
% This implies that large-sized transfers are very likely to be \kvcache.
%
At misprediction, the applications may process the data before decryption.
To detect usage-before-decryption,
\sys revokes the read and write permissions of the application for
the plaintext which is a placeholder before decryption finishes.
%based on the observation that the CPU does not alter the \kvcache,
If the application attempts to
access (read or write) this data, a page fault occurs.
\sys then decrypts the data synchronously and allows the application to
continue its execution.

\section{Implementation}

% \subsection{Side-Channal-Free Ciphertext Placement}
% \label{subsec:design-allocation}

% \TODO{Divide into shared memory management and private memory management}

Current CUDA uses a zero-copy technique for ciphertext in CVM shared memory:
for example, on host-to-device memory copy,
the plaintext is stored in CVM private memory before encryption;
the encryption loads the plaintext and directly writes the ciphertext in CVM shared memory.
In our design, employing a zero-copy technique could expose uncommitted ciphertext, posing a potential security risk.
To mitigate this,
our system (\sys) stores the predicted ciphertext in CVM private memory and only transfers it to shared memory after validating the prediction.
Therefore, unvalidated ciphertext is not exposed in CVM shared memory to attackers and the security is not compromised.
To eliminate the memory copy overhead,
\sys leverages fix-sized buffers in shared memory for DMA to eliminate memory allocation from %\yifan{from?}
the critical path
and pipelines data copy with PCIe transfer.
Because memory copy is faster than PCIe, the pipeline does not require much buffer for the memory copy stage,
minimizing the usage of CVM shared memory.

\iffalse
\sys limits the private memory usage by an encryption window.
The window size is configured based on application behaviors,
for example, when facing model offloading, the window is set to be the size of the next layer.
The memory of ciphertext in the window would not be released until the prediction is commited, or dropped due to prediction errors.
\fi

We implement a prototype of \sys as user-space runtime.
We hack CUDA APIs related to CPU-GPU data transfer (e.g., \texttt{cudaMemcpyAsync})
to implement the pipelining.
Although CUDA implements the coupled memory copy API in proprietary code,
it calls OpenSSL APIs (\texttt{EVP\_EncryptUpdate}) encryption and decryption.
\sys also hacks those OpenSSL APIs to decouple encryption or decryption from the memory copy API.
To decouple encryption from host-to-device copy,
\sys copies the pre-encrypted ciphertext to the destination buffer provided by CUDA.
For decryption, the corresponding API called by CUDA would return without performing the decryption.
which is pipelined later.
% A copy is required because the bounce buffer would be used for other data transfers.
% The copy latency of both cases is hidden with PCIe transfer.
\sys uses
MPK/PKU~\cite{intel-mpk} to implement the read/write access revoke with little overhead,
which can apply to Intel TDX and AMD SEV-SNP.
The core logic of \sys consists of 958 lines of C++ code,
with another about 600 lines of code for hacking the APIs.
% While the swapping of the evaluated applications only happens in user mode,
% data transfer of other inference system can also happen in kernel-plane NVIDIA driver,
% and we believe \sys can be easily ported to kernel mode.

\section{Evaluation}
\label{sec:eval}

% \TODO{The performance of {\sys} requires retesting.}
% \TODO{Add figures}
We answer the following questions in the evaluation section:
\begin{enumerate}   
    \item What degree of performance improvement does {\sys} bring to LLM inference and
        fine-tuning %\yifan{fine-tuning?}
        systems in \cc?
    \item To what extent does the pipelining contribute to the overall performance improvement?
    % \item What are the primary factors contributing to the overhead in {\sys}?
    \item Does \sys's performance rely on the high accuracy of predictions? %rate of \kvcache swapping?
    % \item How much memory does {\sys} consume?
\end{enumerate}

% We evaluate {\sys} with windowed ciphertext placement as default.
% The inplace ciphertext plaintext policy is just evaluated in memory usage.

\subsection{Evaluation Setup}
\label{subsec:eval-setup}

\heading{Platform configurations.}
We setup a KVM VM on an Intel server equipped with dual Intel Xeon Platinum 8462Y+ CPUs at 4.10 GHz.
This VM is configured with 16 virtual CPUs (vCPUs) and 250GB of memory, running Ubuntu 20.04 LTS.
Additionally, a single H100-SXM GPU, connected via a PCIe 5.0 link capable of a maximum duplex bandwidth of 128GB/s,
is integrated into the server. This GPU is directly passed through to the VM for dedicated use.

\heading{Workloads and metrics.}
We conducted experiments on the cutting-edge inference systems, FlexGen~\cite{flexgen-paper},
PEFT~\cite{peft} and vLLM~\cite{vllm-paper}, under \cc environment.
These LLM applications utilize model offloading and \kvcache swapping to address GPU memory shortage.
FlexGen is a leading LLM inference system renowned for its model swapping capabilities in high-throughput scenarios.
PEFT is the state-of-the-art LLM training system with support of LoRA fine-tuning and integration of DeepSpeed~\cite{deepspeed-zero} model offloading.
vLLM, the state-of-the-art LLM inference system in serving scenarios,
uses \kvcache swapping to handle the GPU memory shortage.
% Our experimental configurations closely align with those described in their respective papers.
We utilize FlexGen to assess the performance of model offloading, with synthetic datasets to evaluate inference throughput.
In each test case of FlexGen, 1000 requests are generated and we report the average throughput.
We run PEFT fine-tuning using ultrachat~\cite{ultrachat-dataset} dataset for one epoch (about 6k sequences) and report the training throughput.
Additionally, we used vLLM to evaluate performance in KV cache swapping, utilizing ShareGPT~\cite{sharegpt-web} and Alpaca~\cite{alpaca-web} to cover a range of request lengths.
vLLM's focus is on serving, hence we evaluated normalized latency across varying request rates to demonstrate its serving capabilities.
In each test case of vLLM, 30-minute traces are used to evaluate the systems.
% Furthermore, we tested FlexGen with both model offloading and KV cache swapping to measure throughput in a comprehensive scenario.

\heading{Baselines.}
We conducted a comparative analysis of \sys against two baseline systems.
The first baseline, referred to as ``w/o CC'' (without Confidential Computing),
represents the native performance of inference systems,
without the confidential computing feature.
The second baseline, denoted as ``CC'' (Confidential Computing),
showcases the performance of the current NVIDIA Confidential Computing framework specifically applied in LLM inference.

\begin{figure}
    \subfloat[FlexGen OPT-66B] {
        \includegraphics[width=0.33\linewidth]{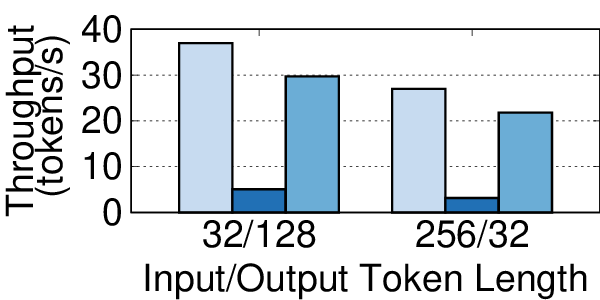}
        \label{fig:eval-flexgen-opt-66b}
    }
    \subfloat[FlexGen OPT-175B] {
        \includegraphics[width=0.33\linewidth]{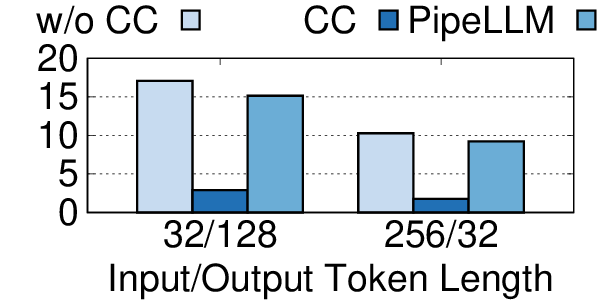}
        \label{fig:eval-flexgen-opt-175b}
    }
    \subfloat[PEFT] {
        \includegraphics[width=0.33\linewidth]{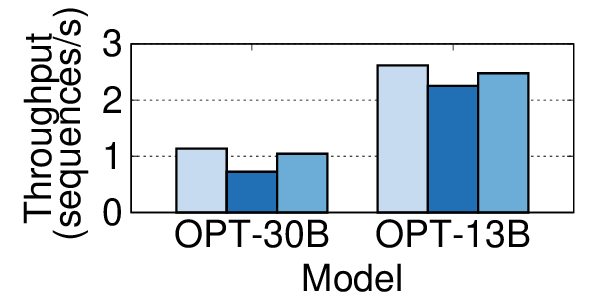}
        \label{fig:eval-peft}
    }
    \caption{\textbf{Performance of FlexGen and PEFT with model offloading.}}
    % \textmd{X-axis is configuration of FlexGen.}
    \label{fig:eval-flexgen}
    \vspace{-4ex}
\end{figure}

\begin{figure*}
    \subfloat[Alpaca parallel size = 2]{
        \includegraphics[width=0.33\linewidth]{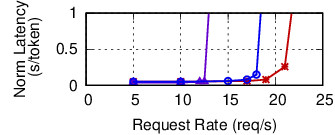}
        \label{subfig:eval-vllm-opt-30b-alpaca-2}
    }
    \subfloat[Alpaca parallel size = 4]{
        \includegraphics[width=0.33\linewidth]{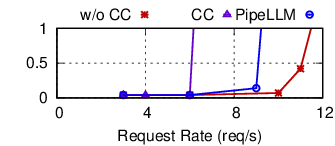}
        \label{subfig:eval-vllm-opt-30b-alpaca-4}
    }
    \subfloat[Alpaca parallel size = 6]{
        \includegraphics[width=0.33\linewidth]{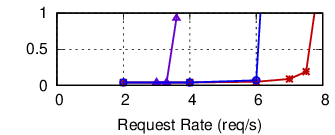}
        \label{subfig:eval-vllm-opt-30b-alpaca-6}
    }
    \hfill
    \subfloat[ShareGPT parallel size = 2]{
        \includegraphics[width=0.33\linewidth]{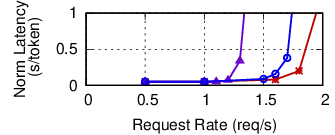}
        \label{subfig:eval-vllm-opt-30b-sharegpt-2}
    }
    \subfloat[ShareGPT parallel size = 4]{
        \includegraphics[width=0.33\linewidth]{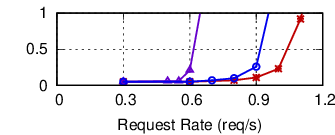}
        \label{subfig:eval-vllm-opt-30b-sharegpt-4}
    }
    \subfloat[ShareGPT parallel size = 6]{
        \includegraphics[width=0.33\linewidth]{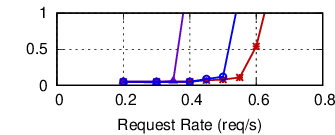}
        \label{subfig:eval-vllm-opt-30b-sharegpt-6}
    }
    \vspace{1ex}
    \caption{\textbf{Performance on vLLM OPT-30B model with \kvcache swapping.}}
    % \textmd{\TODO{More configurations}}
    \label{fig:eval-vllm-opt-30b}
\end{figure*}

\subsection{End-to-End Performance}
\label{subsec:eval-perf}

We demonstrate that {\sys} effectively reduces overhead in various swapping scenarios, encompassing model offloading and \kvcache swapping.
Remarkably, this reduction in overhead is achieved without compromising security.
Furthermore, the performance enhancements offered by {\sys} are not confined to a particular model or configuration;
they are broadly applicable, underscoring the system's versatility and effectiveness across different setups.

%\subsubsection{Model Offloading}
\heading{Model offloading.}
\label{subsubsec:eval-perf-model}
We conducted tests on model offloading using FlexGen as inference cases and PEFT\cite{peft} as fine-tuning cases.
For FlexGen, our evaluation focused on two specific models: OPT-66B and a 4-bit-quantized version of OPT-175B.
We use multiple configurations for running FlexGen.
The input token length was set to either 32 or 256, and the output was set to either 128 or 32.
For PEFT, OPT-30B and OPT-13B are used for LoRA-based fine-tuning.
The memory requirements of these models exceed the capacity of a single GPU.
To evaluate model offloading, we maintained the \kvcache and runtime temporary data on the GPU, limiting the offloaded data exclusively to model weights.

A key aspect to note is that the performance of model offloading is bound by PCIe throughput.
For example, the peak swapping throughput of ``w/o CC'' in FlexGen is about 56GB/s.
To maintain optimal performance, it is imperative for \sys to generate ciphertext at a speed that matches or exceeds the swapping throughput. 
Therefore, \sys would utilize multiple CPU threads dedicated to encryption and decryption to optimize model offloading.

% In test cases of PEFT, \nvcc has a performance of 22.07 second per iteration, compared with non-CC's 14.06 second per iteration, a 36.4\% performance drop. \sys's performance is 17.2 second per iteration, lowering the overhead to 18.25\%.

Figure \ref{fig:eval-flexgen} presents the \sys's performance in running FlexGen and PEFT in the context of model swapping with various models and input configurations.
Enabling confidential computing typically results in a substantial performance decline, ranging from 82.8\% to 88.2\% in FlexGen,
and up to 36.2\% in PEFT.
%This decrease is a notable concern, as it significantly hampers the efficiency of the system.

\sys successfully mitigates this performance degradation, reducing the overhead to less than 19.6\%.
It is a great improvement and highlights the efficacy of \sys in maintaining system performance even under the constraints of confidential computing.
The remaining overhead mainly owes to 40GB/s maximum bandwidth of CPU-to-GPU memory copy
even if all overhead of CPU-side encryption is eliminated.
%The remaining overhead mainly comes from the first iteration of the inference,
%with no knowledge on the model architecture and the prediction does not work.
% The ability of \sys to achieve such a low overhead while still adhering to the stringent security protocols of confidential computing is noteworthy.
% It not only enhances the overall efficiency of the model swapping process but also ensures that the security integrity of the system is uncompromised.
% This balance between performance and security is crucial in environments where data sensitivity is paramount, making \sys an invaluable asset in such settings.

%\subsubsection{K/V Cache Swapping}
\heading{\kvcache swapping.}
\label{subsubsec:eval-perf-kv}
We conducted tests with vLLM, the state-of-the-art LLM inference system in serving scenarios.
Our evaluation focused on two models: OPT-30B and OPT-13B.
During these evaluations, we ensured that all model weights for these models were stored in the GPU. Our experimental approach involved multiple configurations.
We set the parameter of parallel sampling to 2, 4, and 6, in consistency with the methodologies outlined in the original paper.
For vLLM, \sys only uses one thread for encryption and one thread for decryption.

Figure \ref{fig:eval-vllm-opt-30b} provides a comprehensive analysis of {\sys}'s performance in handling \kvcache swapping for the OPT-30B model.
The data reveals that \cc's encryption and decryption results in an overall performance drop, ranging from 33.3\% to 52.8\%.
{\sys} reduces the overhead to 5.2\% - 14.2\%.
Although the prediction success rate is near 100\%,
\sys is not able to reduce the overhead of decreased throughput of data transfer (from 64GB/s to about 40GB/s)
as well as the control-plane overhead of \nvcc  (as analyzed in \S\ref{sec:analysis}).
% the time between swap in and swap out of some data may not be enough for decryption and encryption.
% The remaining overhead mainly comes from this.

% \begin{figure*}
%     \subfloat[opt-13b alpaca best-of-2]{
%         \includegraphics[width=0.33\linewidth]{./figures/eval-vllm-opt-13b-alpaca-2.eps}
%         \label{subfig:eval-vllm-opt-13b-alpaca-2}
%     }
%     \subfloat[opt-13b alpaca best-of-4]{{\sys} 
%     \caption{\textbf{K/V Cache Swapping for opt-13b model}}
%     \textmd{\TODO{This figure is not useful.}}
%     \label{fig:eval-vllm-opt-13b}
% \end{figure*}

% Figure \ref{fig:eval-vllm-opt-13b} shows the performance of {\sys} in K/V cache swapping of opt-13b model.
% \sys, \cc and the baseline have similar performance in opt-13b model.
The observed performance degradation of ``CC'' with the OPT-13B model in ShareGPT dataset is 15.3\% - 23.6\%,
and less than 8\% overhead in Alpaca dataset, notably lower than that experienced with the OPT-30B model,
owing to the distinct memory usage characteristics of the two models.
Specifically, OPT-13B's model weights occupy only about 26GB, which is 32.5\% of the GPU memory, in contrast to OPT-30B, which utilizes a significantly larger portion, approximately 60GB or 75\% of the GPU memory.
However, \sys can still reduce the overhead to less than 8\% in OPT-13B cases.

\iffalse
\subsubsection{Mixed Swapping}
\begin{figure}
    \subfloat[OPT-66B model] {
        \includegraphics[width=0.5\linewidth]{./figures/eval-flexgen-opt-66b-mixed.eps}
        \label{fig:eval-flexgen-opt-66b-mixed}
    }
    \subfloat[OPT-175B model] {
        \includegraphics[width=0.5\linewidth]{./figures/eval-flexgen-opt-175b-mixed.eps}
        \label{fig:eval-flexgen-opt-175b-mixed}
    }
    \vspace{1ex}
    \caption{\textbf{Performance of FlexGen with mixed swapping.}}
    % \textmd{X-axis is configuration of FlexGen.}
    \label{fig:eval-flexgen-mixed}
\end{figure}

We executed FlexGen to evaluate its performance in mixed swapping scenarios, with both model offloading and \kvcache swapping enabled.
This configuration is akin to the one described in Section \ref{subsubsec:eval-perf-model}. However, it differs in that it involves swapping both the model weights and the K/V cache.

Figure \ref{fig:eval-flexgen-mixed} presents the performance results of {\sys} during mixed swapping. The introduction of confidential computing results in a significant performance reduction, ranging from 86.3\% to 91.9\%.
However, the deployment of {\sys} notably mitigates this overhead, reducing it to approximately 3.7\%.
\fi

\subsection{Analysis of Pipelining}

% This subsection answers the following question:

% \begin{itemize}
%     \item How does {\prediction} contribute to the performance?
%     \item How does prediction failure handling (NOP insertion) contribute to the performance?
% \end{itemize}

% To evaluate the contribution of {\prediction}, we evaluate method with multi-thread only.

% This subsection aims at evaluating the performance contribution of {\prediction} in \ref{subsec:design-prediction} and NOP insertion in \ref{subsec:design-prediction-failure}.

% To evaluate {\prediction}, we add another baseline which only ultilizes multi-thread encryption without prediction, naming \textit{w/o prediction}.
% We also test the system with {\prediction} only without NOP insertion, named \textit{w/ prediction}.

% \noindent\textbf{Prediction.}

% \begin{figure}
%     \includegraphics[width=\linewidth]{./figures/eval-flexgen-opt-66b-bd.eps}
%     \caption{Compared with trivial multi threading on FlexGen.}
%     \label{fig:eval-flexgen-opt-66b-bd}
% \end{figure}

% In model swapping, we compare {\sys} with trivial multi-threading without prediction.

% \yifan{FlexGen does not leave much space for overlap. The bottleneck is PCIe.
% So I do not plan to add FlexGen in this part.}

\sys introduces pipelining to enhance performance, utilizing multi-threading to boost encryption throughput.
It is important to note that ``CC'' can also use multiple CPU threads to attain a line-rate encryption rate on the CPU, thereby narrowing the performance gap.
Our findings demonstrate that ``CC'' requires a greater number of threads compared to \sys to achieve comparable performance.

\begin{figure}
    \includegraphics[width=\linewidth]{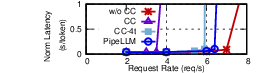}
    \vspace{1ex}
    \caption{\textbf{Performance of trivial multi threading on vLLM OPT-30B.} ``CC-4t'' refers to the system using 4 threads for encryption and decryption without pipelining.}
    \label{fig:eval-vllm-opt-30b-bd}
\end{figure}

We conducted the experiment using the vLLM with the OPT-30B model.
The dataset employed was Alpaca, and the parallel size set for the experiment was 6.
Figure \ref{fig:eval-vllm-opt-30b-bd} illustrates the performance comparison between ``CC'', utilizing four threads for encryption and decryption, and the alternatives 'w/o CC' and \sys.
Notably, \sys only uses two threads and yet outperforms ``CC'' with four threads but in the absence of pipelining.
This highlights the effectiveness of pipelining in this context.

\label{ss:wrongpredicrion}
\subsection{Ablation Study on Success Rate}
\sys achieves near 100\% success rate on \kvcache swapping in vLLM,
because vLLM takes LIFO as its swap policy, bringing benefit to \sys.
To show that \sys works well on inference systems with other swap policies on \kvcache,
we conduct experiments on prediction failure on the sequence.

\begin{figure}
    \includegraphics[width=\linewidth]{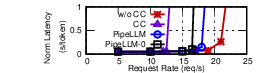}
    \vspace{1ex}
    \caption{\textbf{Ablation study on success rate.} ``\sys-0'' refers to the system with zero success rate on sequence prediction.}
    \label{fig:eval-vllm-opt-30b-alpaca-2-hit}
\end{figure}

We use vLLM with the OPT-30B model, Alpaca dataset, and a parallel size of 2 as an example.
Figure \ref{fig:eval-vllm-opt-30b-alpaca-2-hit} shows that the performance of \sys with zero prediction success rate only slightly drops by 8.3\%, mainly caused by the overhead of NOPs.
%Low success rate would only
Upon sequence prediction failure, \sys can still use the ready ciphertext and use NOP to drop the mispredicted ciphertext.
The additional encryption latency is well hidden with GPU computation.

\iffalse
\subsection{Memory Usage}
This subsection evaluates the memory overhead of {\sys}.
FlexGen running on ``w/o CC'' consumes 240GB memory, while \sys uses less than 4GB additional memory.
vLLM consumes 97GB memory on ``w/o CC'', including 60GB memory reserved for \kvcache swapping.
\sys with windowed ciphertext placement uses 1GB to 30GB more memory in vLLM.
\fi

% \noindent\textbf{Comparison of {\zerocopy} and {\inplace}.}
% We use vLLM to compare {\zerocopy} and {\inplace}.
% Figure \TODO{} shows the memory usage of the two {\sys} using the two policies during the test.
% {\inplace} saves the memory usage, with no performance drop.

% \begin{figure}
%     \includegraphics[width=\linewidth]{./figures/eval-vllm-opt-30b-sharegpt-4-memory.eps}
%     \caption{Overhead caused by limited encryption window.}
% \end{figure}
% We use vLLM to evaluate the benefit brought by {\inplace}.
% We compare {\sys} with the system using {\zerocopy}, adding different memory usage bound.
% If encryption window is set to 5GB, the performance would drop 9\%.

% Simulation platform
\section{Discussion}
\label{sec:discussion}

% \TODO{Write this section after finishing evaluation part}

\iffalse
\subsection{Evaluation Platform}
\label{ss:simulationdetails}
Owing to ecosystem constraints, we do not have a full-fledged H100 \cc platform.
Additionally, NVIDIA's \cc software is incompatible without a CC-enabled H100.
Therefore, we resorted to using available H100 hardware to emulate the confidential computing platform.
Here we explain the difference between actual CC platform and our emulated platform
demonstrating that these differences do not significantly impact our results.

\noindent\textbf{The simulation of H100 confidential computing.}
We simulated the CPU-side encryption and decryption procedures,
identified as the primary sources of performance overhead in NVIDIA's confidential computing architecture\cite{h100cc-paper}.
Considering that other modifications in NVIDIA's CC architecture do not significantly influence performance,
we choose not to include them in our simulation.

\noindent\textbf{Using traditional VM to emulate CVM.}
In the context of LLM inference, the overhead associated with CVM hardware is not significantly higher compared to traditional VM setups.
CPU execution within CVMs incurs approximately a 4\% overhead\cite{sevsnp-perf}.
Since the major I/O processes are GPU-related, and these overheads have been emulated in this study,
using a traditional VM to simulate a CVM is likely to yield minimal performance discrepancies.
% While VM exits are costly \cite{li2023bifrost},
% the CPU hardware feature of direct interrupt injection prevents most of 
\fi

\subsection{Security Analysis}
\label{ss:securityanalysis}

% The \zerocopy explained in \S\ref{subsec:design-allocation} would bring cache attack \cite{Lipp2018meltdown, Kocher2018spectre},
% which is a form of side channel attack.
% CPU cache in range of CVM shared memory is not protected.
% Therefore, directly calculating on CVM shared memory would expose cache information to attackers,
% potentially leaking sensitive data.
% \sys avoids using \zerocopy policy in its design.

% \sys does not introduce additional attack surfaces in comparison to \nvcc.
% Crucially, in the design of \sys, no cryptographic data is transferred to the shared memory until a prediction is verified.
% The confidentiality and integrity of this ciphertext are safeguarded by the AES-GCM encryption algorithm.
% Furthermore, the padding NOPs, as outlined in Section \S~\ref{subsec:nop-insertion}, contains rubbish data to prevent leaking useful data to attackers.
% As a result, the ciphertext of any mis-predicted transfer does not pose additional security risks.

% \TODO{Nvidia currently supports single GPU enclave only. So we focus on single GPU use cases
% (leave multi-GPU cases for future work)}

% Confidentiality and integrity
\sys preserves the confidentiality and integrity features of \nvcc.
The MPK/PKU mechanism used for validating predictions in \sys operates within
the guest VM and is secured from attackers by CVM hardware.
Importantly, in \sys's design, no cryptographic data is
transferred to shared memory until a prediction is verified.
Additionally, the padding NOPs described in \S\ref{subsec:nop-insertion} contain dummy data,
preventing leakage of information to attackers.
Consequently, the ciphertext of any mispredicted transfer does not compromise
the security guarantees of \nvcc.

\label{ss:sidechannel}
% Side channel
Although \sys does not compromise confidentiality or integrity,
its mis-speculation introduces side channels in NOP transfers compared to \nvcc, including
(1) attackers can detect if the LLM system is currently swapping by observing NOPs,
and (2) attackers could profile the frequency of prediction failures,
potentially revealing the swapping patterns of applications.
% \TODO{Just list the side channels}
% However, we argue that the side channel is insignificant
% because the information about swapping could be obtained from existing side channels
% (e.g., data sizes) which are not involved in the threat model of this paper.
The security implications of these side channels remain unclear and need further research.

\subsection{Should Swap Data Be Re-encrypted?}
\label{ss:re-encryption}

As discussed in \S\ref{sec:analysis}, the primary overhead of \nvcc in
LLM workloads arises from encrypting swap data.
Observing that applications on the CPU do not modify this data,
one possible approach is to retain the encrypted version on the CPU.
This would allow the encrypted data to be transferred directly to the GPU
during loading, thereby eliminating encryption overhead.
While this approach improves performance, implementing it naively compromises security guarantees.
For example, reusing encrypted data enables attackers to identify data that matches a
previous transfer; more critically, it could make the system vulnerable to
replay attacks~\cite{h100cc-whitepaper}.

Currently, \nvcc uses the AES-GCM algorithm with incrementing IVs
to prevent ciphertext reuse and defend against replay attacks (\S~\ref{subsec:bg-h100}).
This approach prioritizes simplicity and transparency at the cost of
performance: it encrypts read-only data on-the-fly, incurring some overheads,
but it requires no modifications to applications or encryption engines.
\sys follows this design choice,
providing practical performance benefits while preserving transparency for users.
%\yifan{, (followed by reviewers' guide in the email)}.

However, this design is a trade-off rather than a fundamental limitation.
Future confidential GPUs could benefit from dedicated hardware and
library interfaces to support ciphertext reuse for swap data, potentially
alleviating the bottlenecks faced by today's \ccenabled LLM systems.

%\sys is able to achieve low overhead GPU confidential computing on legacy code that is not specialized for CC.

\subsection{Compared with TEE I/O Approach}
\label{ss:adaptation}
%By the time of paper submission,
%CVM requires CPU cores to encrypt I/O data.

The overhead of \ccenabled LLMs could also by mitigated by hardware encryption.
The next generation of CVM introduces TEE I/O~\cite{tee-io},
equipped with dedicated hardware on CPU SOC for line-rate encryption.
However, limited performance information is currently available.
In practice, a GPU server typically runs multiple VMs---a standard
H100 server, for example, has two CPUs (dual-socket) and eight GPUs---raising
questions about whether the TEE I/O hardware can sustain GPUs' throughputs.
Compared to hardware solutions, \sys offers greater flexibility
for different hardware configurations.

%To enable line-rate encryption, the encryption throughput
%per CPU should be at least four times compared as that of one GPU,
%making hardware approach much more costly and less flexible.

\section{Related Work}
\label{sec:related}

\heading{Enclaves that target CPU}
Various commercial products have been released to provide CPU enclaves for 
protecting user data and code from untrusted operating systems and hypervisors.
An enclave has isolated CPU states and memory region protected by the hardware,
so that the CPU and memory data cannot be directly 
accessed by the privileged software.
Intel SGX~\cite{baumann2015shielding} is the first enclave implementation that provides a process-level enclave,
which mandates the application to be modified to use the SGX interface.

The CVM abstraction, including AMD SEV~\cite{sevsnp, seves}, Intel TDX~\cite{tdxabispec, tdxwhitepaper} and ARM CCA~\cite{armcca},
enables a more application-transparent approach by 
running the user workloads in VM user mode without modifications~\cite{li2021twinvisor, chen2023security, cloudvisor-d, cpc-atc24,li2023bifrost}.
However, for the machine learning workloads, especially LLMs,
these CPU-side enclaves should be combined with GPU-side enclave to provide 
end-to-end protection for them.
%While this work targets CVM+H100 platform, it could apply on other platforms.

\heading{GPU Enclave}
In addition to software-based GPU enclaves,
researchers have also explored hardware-based approaches.
Graviton~\cite{volos2018graviton} represents the pioneering effort in creating a GPU enclave,
achieved by modifying the GPU's command processor and using data encryption to safeguard data confidentiality.
HIX~\cite{jang2019heterogeneous}, without altering GPU hardware,
proposes modest extensions to the I/O interconnect and the memory management unit (MMU)
to secure GPU computations.
Another innovative approach, HETEE~\cite{zhu2020enabling},
leverages PCIe switch fabric to control access to GPUs securely and flexibly without necessitating hardware modifications.
StrongBox~\cite{deng2022strongbox} introduces an integrated-GPU enclave for ARM devices,
utilizing the pre-existing TrustZone features.
Although this paper focuses specifically on the design of \sys for the NVIDIA H100 GPU,
the design considerations are equally applicable to other GPU enclaves.

% [cheng: other possible related topics]
%\heading{Verifiable ML.}

\heading{GPU Attacks.}
Several studies have scrutinized the exploitation of security vulnerabilities to compromise GPU isolation.
A notable vulnerability involves the potential breach of GPU security,
allowing for the execution of GPU-based malware.
This malware can circumvent IOMMU protections and illicitly access private information from the host CPU~\cite{zhu2017understanding}.
Another common avenue of attack targets residual memory contents left on the GPU from previous computations.
For instance, Lee et al.\cite{lee2014stealing} demonstrated methods of extracting sensitive webpage data retained in GPU memory,
particularly from Chromium and Firefox browsers.
Similarly, CUDA Leaks\cite{pietro2016cuda} revealed the possibility of extracting sensitive information,
including plaintext and encryption keys, from the GPU's shared and global memory.

\heading{LLM-Specific Optimizations.}
Extensive research has been conducted to enhance the performance of LLM without CC,
such as operator-level optimizations~\cite{flashattention,flashattention2},
efficient memory management on GPU~\cite{vllm-paper,sheng2024slora},
and optimizations based on sparsity~\cite{powerinfer,powerinfer-2,turbosparse}.
\sys is orthogonal to these optimizations if running within CC.

\section{Conclusion}
\label{sec:concl}

This paper comprehensively evaluates and analyzes the performance overhead when serving and fine-tuning
LLMs on CVM equipped with a GPU enclave.
This paper introduces \sys, a user-transparent runtime system that
effectively reduces the performance overhead associated with GPU-based confidential computing.
%By utilizing techniques inspired by CPU instruction pipelining,
\sys uses speculative pipelined encryption to overlap encryption and data transmission,
thereby significantly minimizing latency.
Our performance evaluations with real-world LLM systems,
vLLM, PEFT, and FlexGen, demonstrate that \sys
can reduce overhead to $<$19.6\% across various LLM sizes.

\section*{Acknowledgments}
\label{sec:ack}

We express our sincere gratitude to our shepherd Kiwan Maeng,
and the anonymous reviewers for their insightful comments.
This work was partially supported by NSFC (No. 62372287 and 61925206).
Cheng Tan is supported in part by
NSF CAREER Awards \#2237295.
Zeyu Mi (yzmizeyu@sjtu.edu.cn) is the corresponding author.

%-------------------------------------------------------------------------------
% \bibliographystyle{plain}
% \bibliography{\jobname}

\balance{
\small{
\bibliographystyle{plain}
\bibliography{references}
}
}

%%%%%%%%%%%%%%%%%%%%%%%%%%%%%%%%%%%%%%%%%%%%%%%%%%%%%%%%%%%%%%%%%%%%%%%%%%%%%%%%
\end{document}